\documentclass[12pt]{article}

\usepackage{amsmath,amssymb,amsthm}
\usepackage{graphicx,psfrag,epsf}
\usepackage{hyperref}
\usepackage{enumerate}
\usepackage[authoryear,round]{natbib}
\usepackage{chngcntr}
\usepackage{titlesec}
\usepackage{thmtools}
\usepackage{thm-restate}
\makeatletter
\theoremstyle{plain}
\declaretheorem[name=Theorem]{thm}
\theoremstyle{definition}

\newcommand{\E}{\mbox{$\mathbb{E}$}}
\newcommand{\real}{\mathbb{R}}
\let\hat\widehat
\let\tilde\widetilde

\newcommand{\blind}{1}

\newenvironment{enum}{
\begin{enumerate}
  \setlength{\itemsep}{1pt}
  \setlength{\parskip}{0pt}
  \setlength{\parsep}{0pt}
}{\end{enumerate}}

\usepackage[utf8]{inputenc}
\usepackage{lscape}
\usepackage{tablefootnote}
\usepackage{colortbl}
\usepackage{caption}
\usepackage{float}
\usepackage{multirow}
\usepackage{subcaption}
\usepackage{float}
\usepackage{changepage}
\usepackage{algorithm}
\usepackage{diagbox}
\usepackage[noend]{algpseudocode}
\makeatletter
\def\BState{\State\hskip-\ALG@thistlm}
\makeatother

\newcommand{\double}{\text{dbl}}
\newcommand{\poolCDF}{\text{poolCDF}}
\newcommand{\sub}{\text{sub}}
\newcommand{\rep}{\text{rep}}
\newcommand{\modelAvg}{\text{modAvg}}
\newcommand{\isolate}{\text{isolate}}
\newcommand{\shrinkage}{\text{shrinkage}}
\newcommand{\probnew}{\overline{\Pi}}
\newcommand{\probnewaug}{\overline{\Pi}^{\text{aug}}}
\newcommand{\probnewone}{\overline{\Pi}^{(1)}}

\usepackage{secdot}
\usepackage{sectsty}
\sectionfont{\centering\normalsize}
\subsectionfont{\normalsize}

\makeatletter
\renewcommand{\@seccntformat}[1]{%
  \ifcsname format@#1\endcsname
    \csname format@#1\endcsname
  \else
    \csname the#1\endcsname\quad % the default
  \fi
}
\g@addto@macro\appendix{%
  \def\format@section{APPENDIX \thesection: }%
}
\makeatother

\begin{document}

\def\spacingset#1{\renewcommand{\baselinestretch}%
{#1}\small\normalsize} \spacingset{1}

%%%%%%%%%%%%%%%%%%%%%%%%%%%%%%%%%%%%%%%%%%%%%%%%%%%%%%%%%%%%%%%%%%%%%%%%%%%%%%

\if1\blind
{
  \title{\bf Distribution-Free Prediction Sets for\\\vspace{.2cm} Two-Layer Hierarchical Models}
  \author{Robin Dunn$^1$\thanks{
    Robin Dunn is a Principal Statistical Consultant at Novartis Pharmaceuticals Corporation (e-mail: robin.dunn@novartis.com). Larry Wasserman is a Professor in the Department of Statistics \& Data Science and the Machine Learning Department, Carnegie Mellon University (e-mail: larry@stat.cmu.edu). Aaditya Ramdas is an Assistant Professor in the Department of Statistics \& Data Science and the Machine Learning Department, Carnegie Mellon University (e-mail: aramdas@stat.cmu.edu).}, Larry Wasserman$^{2,3}$, Aaditya Ramdas$^{2,3}$\hspace{.2cm}}
   \date{ \small
    $^1$Novartis Pharmaceuticals Corporation, Advanced Methodology and Data Science, East Hanover, NJ USA \\
    $^2$Department of Statistics \& Data Science, Carnegie Mellon University, Pittsburgh, PA USA \\
    $^3$Machine Learning Department, Carnegie Mellon University, Pittsburgh, PA USA
   }
  \maketitle
} \fi

\if0\blind
{
  \bigskip
  \bigskip
  \bigskip
  \begin{center}
    {\LARGE\bf Distribution-Free Prediction Sets for\\\vspace{.2cm} Two-Layer Hierarchical Models}
\end{center}
  \medskip
} \fi

\bigskip
\begin{abstract}
We consider the problem of constructing 
distribution-free prediction sets
for data from two-layer hierarchical distributions.
For iid data,
prediction sets can be constructed using 
the method of conformal prediction.
The validity of conformal prediction hinges on the exchangeability of the data,
which does not hold when groups of observations come from distinct distributions, such as multiple observations on each patient in a medical database.
We extend conformal methods to a hierarchical setting.
We develop CDF pooling, single subsampling, and repeated subsampling approaches to construct prediction sets in unsupervised and supervised settings. We compare these approaches in terms of coverage and average set size. 
If asymptotic coverage is acceptable, we recommend CDF pooling for its balance between empirical coverage and average set size. If we desire coverage guarantees, then we recommend the repeated subsampling approach.
%Simulations show that this approach has coverage and set sizes close to the single subsampling method, and the repeated subsampling approach is more reproducible.
Supplementary materials are available online.
\end{abstract}

\noindent%
{\it Keywords:}  conformal prediction, random effects, model misspecification, subsampling.
\vfill

\newpage
\spacingset{1.45} % DON'T change the spacing!

\section{INTRODUCTION} \label{sec:intro}

% Let $(X_1,Y_1),\ldots, (X_n,Y_n)$ be $n$ iid pairs of observations from a distribution $P$.
Let $Y_1, Y_2, \ldots, Y_n$ be $n$ independent and identically distributed (iid) observations from a distribution $P$.
%Suppose $1-\alpha$ is the user-specified confidence level and $(X,Y)$ denotes a new pair drawn from $P$.
Suppose $1-\alpha$ is the user-specified confidence level and $Y$ denotes a new observation drawn from $P$.
In set-valued, unsupervised prediction, we want to find a 
set-valued function $C$ such that
\begin{equation}\label{eq::required}
P( Y \in C(\alpha)) \geq 1-\alpha.
\end{equation}
(We should really write
$P^{n+1}( Y \in C(\alpha)) \geq 1-\alpha$
since the randomness is over $Y$ and the training data.
We have suppressed the superscript for simplicity.)
\cite{vovk2005algorithmic}
created the method of conformal prediction
to construct $C$ such that
(\ref{eq::required}), or $P(Y \in C(X; \alpha)) \geq 1-\alpha$ in the supervised case,  holds for all distributions $P$.
In other words, conformal methods yield distribution-free
prediction sets.

A fundamental assumption
of the usual conformal method is that the data are iid
(or, at least, exchangeable).
We extend conformal methods
to the following hierarchical model where the iid assumption fails.
Let $P_1,P_2,\ldots, P_k \sim \Pi$ be random distributions
drawn from $\Pi$.
%$${\cal D}_j = \{ (X_{j1},Y_{j1}),\ldots, (X_{jn_j },Y_{j n_j}) \}$$
In the unsupervised setting, let
${\cal D}_j = \{ Y_{j1}, Y_{j2}, \ldots, ,Y_{j n_j} \}$
be $n_j$ iid observations drawn from
$P_j$
for $j=1,\ldots, k$.
It is helpful to imagine that we have $k$ subjects
and ${\cal D}_j$ represents $n_j$ observations on subject $j$. 
We know the identity of the group (from 1 to $k$) to which each observation belongs.
We assume that the values of $n_j$ are fixed prior to data collection.

There are two tasks to consider:
\begin{enum}
\item Task 1: Predicting an observation on a new subject.
% Let $P_{k+1}\sim \Pi$ denote a new draw from $\Pi$ (a new subject) and let $(X,Y)\sim P_{k+1}$. The goal is to construct a prediction set for $Y$ using $X$ and the training data ${\cal D}_1,\ldots, {\cal D}_k$.
Let $P_{k+1}\sim \Pi$ denote a new draw from $\Pi$ (a new subject) and let $Y\sim P_{k+1}$. The goal is to construct a prediction set for $Y$ using the training data ${\cal D}_1,\ldots, {\cal D}_k$.
\item Task 2: Predicting a new observation on one of the current subjects.
% Let $(X,Y)$ denote a new draw from one of the distributions $P_j, 1\leq j \leq k$. We want a prediction for $Y$ based on $X$ and the training data.
Let $Y$ denote a new draw from one of the distributions $P_j, 1\leq j \leq k$. We want a prediction set for $Y$ based on the training data.
\end{enum}

In Task 1, the new $Y$ is not exchangeable with the observations from any single observed distribution. In addition, if the training data contains multiple observations from at least one of $P_1, \ldots, P_k$, then the new $Y$ is not exchangeable with the full training data either. Rather than applying standard conformal methods, this setting requires novel approaches that build on the exchangeability of the distributions and the exchangeability of the observations from a given distribution.
By contrast, one valid approach to Task 2 is to construct conformal sets using only the data from the subject of interest. We consider that method, but we also incorporate shrinkage and borrowing strength into a second conformal approach. By leveraging data across subjects, the latter method may produce smaller sets.

We have described these tasks in the unsupervised setting, but we also consider Task 1 in the supervised setting. One example of a supervised two-layer hierarchical model is the random effects working model
$Y_{ij} = \beta_{0j} + \beta_{1j} X_{ij} + \epsilon_{ij}$
where
$\epsilon_{ij}\sim N(0,\sigma^2)$.
Here $P_j$ denotes the true underlying distribution of $(X,Y)$ for group $j$. Suppose this random effects model represents the true relationship between $X$ and $Y$, and suppose $X \sim N(0,1)$. Then drawing $(X_j, Y_j) \sim P_j$ amounts to drawing $X_j \sim N(0,1)$ and $Y_j \sim N(\beta_{0j} + \beta_{1j} X_j, \sigma^2)$. Furthermore, suppose that $P_\beta$ represents the distribution of $(\beta_{0j}, \beta_{1j})$ over the full population. Then drawing $P_j \sim \Pi$ reduces to drawing $(\beta_{0j}, \beta_{1j}) \sim P_\beta$. 
As discussed in Section~\ref{section::basic}, it is possible to use a parametric working model
to get valid prediction sets even if the model is wrong.

\subsection{Related Work}
Key early references on conformal prediction include
\cite{vovk2005algorithmic}
and \cite{shafer2008tutorial}.
The literature on conformal prediction is quickly growing in several overlapping directions. Developments on conformal prediction include connections to traditional statistical methods, extensions to flexible settings, and implementations that are computationally efficient. Work in these directions includes 
interpolations between marginal and conditional coverage \citep{lei2014distribution, barber2021limits},
extensions to multiclass set-valued classification \citep{sadinle2018least}, Mondrian conformal approaches that ensure validity within categories \citep{vovk2005algorithmic}, valid discretizations of conformal methods \citep{chen2018discretized}, anti-conservative bounds on coverage, methods for variable importance, and computationally efficient sample-splitting methods \citep{lei2018distribution}. Many open problems remain in extending conformal methods to new contexts. 

Random effects models are common examples of two-layer hierarchical models. 
%Most of the work on random effects focuses on estimation.
\cite{laird1982random} provide foundational work on the structure and estimation of random effects models for repeated-measures data. The authors note that random effects allow researchers to model both within- and between-subject variation, often using parameters that have natural interpretations. For instance, random effects models frequently are defined by within-subject and across-subject means and variances \citep{dersimonian1986meta}. We incorporate this conceptualization in our simulations. Random effects models have been used for prediction by some researchers in parametric settings
\citep{calvin1991bayesian,
booth1998standard,
schofield2015predictive}. As an alternative to the random effects parametric assumptions, \cite{claggett2014meta} develop methods for inference on the quantiles of study-level parameters without distributional assumptions on these parameters. Thus, researchers have developed some approaches for inference and prediction in random effects parametric settings and for inference on study-specific parametric quantiles without distributional assumptions. 
To the best of our knowledge,
there are no papers
on valid distribution-free prediction for two-layer hierarchical settings.

\subsection{Paper Outline}
Section~\ref{section::basic} reviews conformal prediction. Sections~\ref{section::unsup_methods}, \ref{section::sup_methods}, and \ref{section::obs_methods}  each present methods and simulations for conformal prediction in the two-layer hierarchical setting.
Section~\ref{section::unsup_methods} considers unsupervised prediction on a new distribution. Section~\ref{section::sup_methods} considers supervised prediction on a new distribution. Section~\ref{section::obs_methods}  considers unsupervised prediction on an observed distribution. 
Section~\ref{section::example} implements our supervised prediction methods on data from a sleep deprivation study.
Section~\ref{section::conclusion}
provides concluding remarks.
In the online supplementary material, Appendix~\ref{appendix:math} contains proofs and Appendix~\ref{appendix:sims} contains additional simulations. Code is available at \url{https://github.com/RobinMDunn/ConformalTwoLayer}.

\section{BACKGROUND ON CONFORMAL PREDICTION}
\label{section::basic}

Conformal prediction 
%--- introduced by \cite{vovk2005algorithmic} ---
is a general method for obtaining distribution-free 
prediction sets with confidence guarantees.
Here, we review some background on conformal prediction.

{\bf The Unsupervised Case.}
Let $Y_1,\ldots, Y_n \in {\cal Y}$ be iid observations from a distribution $P$,
and let $Y_{n+1}$ denote a new draw from $P$.
The goal of conformal prediction is to construct a set $C(\alpha)$ based on
the training data
$Y_1,\ldots, Y_n$ such that
$P(Y_{n+1}\in C(\alpha))\geq 1-\alpha$
for every distribution $P$. When $\mathcal{Y} \subseteq \real$ or, more generally, $\mathcal{Y}$ is a linearly ordered set, Theorem~\ref{thm:unsup_basic} provides one valid construction based on order statistics. We say that a method can produce nontrivial sets if $C(\alpha)$ may be a strict subset of $\mathcal{Y}$. 

\begin{restatable}[]{thm}{thmUnsupBasic} \label{thm:unsup_basic}
Define $C(\alpha)=[Y_{(r)}, Y_{(s)}],$ where $r = \lfloor (n+1)(\alpha/2) \rfloor$ and $s = \lceil (n+1)(1-\alpha/2) \rceil$. (If $r < 1$ and $s > n$, set $Y_{(r)} = \min\{\mathcal{Y}\}$ and $Y_{(s)} = \max\{\mathcal{Y}\}$.) Then for every distribution $P$,
$P(Y_{n+1}\in C(\alpha))\geq 1-\alpha$. This method can produce nontrivial sets if $n \geq 2/\alpha - 1$.
\end{restatable}

Theorem~\ref{thm:unsup_basic} relies on the exchangeability of the original sample's order statistics. For a proof, see Appendix~\ref{appendix:math}. Alternative valid constructions rely on the exchangeability of conformal residuals constructed from the sample. For any $u\in {\cal Y} \subseteq \real^d$, let
${\cal A}(u) = (Y_1,\ldots, Y_n,u)$,
which can be thought of as the training data augmented with a guess that $Y_{n+1}=u$.
Define the residual (or nonconformity score)
$R_i(u) = \phi(Y_i, {\cal A}(u))$
where $\phi: \mathcal{Y} \times \mathcal{Y}^{n+1} \to \real$ is any function that is invariant under permutations of
the elements of
${\cal A}(u)$. We wish to test the hypothesis $H_0: Y_{n+1} = u$. The set of all $u$ for which we do not reject $H_0$ at level $1-\alpha$ will provide the $100(1-\alpha)\%$ prediction set. Assuming $Y_{n+1} = u$, we define
\begin{equation}
\pi(u) = \frac{1}{n+1} \sum_{i=1}^{n+1} I( R_i(u) \geq R_{n+1}(u))
\end{equation}
which is the $p$-value for testing
this hypothesis.
Intuitively, the $p$-value for a given $u$ is small if the residuals at most of $Y_1, \ldots, Y_n$ are smaller than the residual at $u$ (i.e., the $p$-value is small if $Y_{n+1} = u$ does not ``conform'' to the original sample).
$\pi(u)$ is a valid $p$-value because under $H_0$, 
$\pi(u)$ follows a super-uniform distribution over $t\in [0, 1]$. 
That is, $P(\pi(u) \leq t) = P(\pi(u) \leq \lfloor t(n+1)\rfloor / (n+1)) \leq \lfloor t(n+1)\rfloor / (n+1) \leq t$. 
Often $P$ is a continuous distribution and $P(\phi(Y_i, \mathcal{A}(u)) = \phi(Y_j, \mathcal{A}(u))) = 0$ for $i\neq j$.
In this case,
$\pi(u)$ is uniformly distributed over the set
$\{1/(n+1), 2/(n+1), \ldots, 1\}$. 
We invert the test to define
$C(\alpha) = \{ u :\ \pi(u) \geq \alpha\}.$

\begin{thm}
\label{thm::basic}
For $C(\alpha)$ as given above,
$P(Y_{n+1}\in C(\alpha))\geq 1-\alpha$
for every distribution $P$. 
For this method to produce nontrivial sets, it must hold that $n > 1/\alpha - 1$.
\end{thm}

See \cite{vovk2005algorithmic} for a proof. 
For the nontrivial condition, note that $\pi(u) \geq 1/(n+1)$ for any $u$. Hence, if $n\leq 1/\alpha - 1$, then $\pi(u) \geq 1/(n+1) \geq \alpha$ for all $u$.
There is great flexibility in the choice of nonconformity score $\phi$.
Every choice leads to a prediction set with valid coverage,
but different choices may lead to smaller sets.
Thus,
the choice of $\phi$ can affect the efficiency of the prediction set
but not its validity; see \cite{lei2013distribution}.

As an example,
let $R_i(u) = |Y_i - \overline{Y}(u)|$
where
$\overline{Y}(u) = (u + \sum_{i=1}^n Y_i)/(n+1)$
is the mean of the augmented data.
Then
$\pi(u) = (n+1)^{-1} \sum_{i=1}^{n+1} I\bigl( |Y_i - \overline{Y}(u)| \geq |u- \overline{Y}(u)|\bigr)$.
Another useful nonconformity score is
$R_i(u) = 1/\hat p_u(Y_i)$
where $\hat p_u$ is a density estimator based on the augmented data.
\cite{lei2013distribution}
showed that this choice is minimax optimal
when some conditions hold. Finally, the density estimator could be based on a working parametric model such as $\mathcal{Q} = (Q_\theta: \theta\in\Theta)$. For example, we could use $R_i(u) = 1/q_{\hat{\theta}(u)}(Y_i)$, where $\hat{\theta}(u)$ is the maximum likelihood estimate based on $(Y_1, \ldots, Y_n, u)$. Importantly, this choice of residual is valid even if $P$ is not in $\mathcal{Q}$.

{\bf The Supervised Case.}
In this case the data are
$(X_1,Y_1),\ldots, (X_n,Y_n) \sim P$.
Let $(X,Y)\sim P$ be a new observation.
We want a set $C(x;\alpha)$
such that
$P(Y\in C(X;\alpha)) \geq 1-\alpha$ for all $P$. 
%In the supervised setting, the conformal set now depends on the $X_i$\,s as well.
As one possibility,
fix $(x,y)$ and
let $\hat m_{(x,y)}$ be a regression estimator
based on the augmented data
$(X_1,Y_1),\ldots, (X_n,Y_n),(X_{n+1},Y_{n+1})$
with
$(X_{n+1},Y_{n+1})=(x,y)$. Where $R_i(x,y) = |Y_i -\hat m_{(x,y)}(X_i)|$,
let
$$
\pi(x,y) = \frac{1}{n+1} \sum_{i=1}^{n+1} I( R_i(x,y) \geq R_{n+1}(x,y)) \\
$$
and
$$
C(x; \alpha) = \Bigl\{y:\ \pi(x,y) \geq \alpha \Bigr\}.\\
$$
Then
$\inf_P P( Y_{n+1}\in C(X_{n+1}; \alpha)) \geq 1-\alpha$.

A second useful choice of conformal residual is
$R_i(x,y) = 1/\hat p(X_i,Y_i)$ where
$\hat p$ is a joint density estimate based on the
augmented data.
See 
\cite{lei2013distribution}
for more details.

{\bf Methods for Each Task.} We have described an order statistic method and a residual method that are valid under minimal assumptions. In the one-dimensional unsupervised case, the order statistic approach is valid if the data are exchangeable. This method is a simpler construction that does not require data augmentation or the choice of a nonconformity score. Alternatively, the residual approach affords more flexibility through the construction of a nonconformity score, and it extends beyond the one-dimensional and unsupervised setting. This method relies on the exchangeability of the residuals, which holds for any permutation-invariant $\phi$ when the underlying data are exchangeable. To construct prediction sets for a new observation on a new subject (Task 1), we use methods based on the original sample's order statistics in the unsupervised setting, and we use the residual method in the supervised setting. To construct prediction sets for a new observation on an existing subject (Task 2), we use the residual method. In the Task 2 setting, the residual method allows us to implement a nonconformity score based on a shrinkage estimator.

\section{UNSUPERVISED PREDICTION FOR A NEW DISTRIBUTION} \label{section::unsup_methods} 

To develop valid prediction sets in the two-layer hierarchical setting, we start with the unsupervised version.
Recall that
the data come in groups
${\cal D}_1, \ldots, {\cal D}_k$
and each group has iid data
${\cal D}_j = \{ Y_{j1},\ldots, Y_{j n_j}\} \sim P_j,$
where $Y_{j1}, \ldots, Y_{jn_j} \in \mathcal{Y} \subseteq \real$ and
$P_1,\ldots, P_k \sim \Pi$. 
More generally, if $\mathcal{Y}$ is a linearly ordered set, the unsupervised methods that do not require continuous CDFs will still hold. (These are Methods 0, 2, and 3, which we will describe in this section.)
Assuming a new distribution $P_{k+1}\sim \Pi$ and $Y\sim P_{k+1}$,
we want a prediction region for $Y$. 
We can construct prediction sets $C(\alpha)$ such that $Y$ is contained in $C(\alpha)$ with probability at least $1-\alpha$, over the randomness in the initial sample and $Y \sim \tilde{\Pi}$, where $\tilde{\Pi} = \int P d\Pi(P)$. More formally, for $y\in\mathcal{Y}$ and $y_j \in \mathcal{Y}^{n_j}$, we define the distribution over these sources of randomness as $$\overline{\Pi}(y, y_1, y_2, \ldots, y_k) = \left\{ \int P(Y \leq y) d\Pi(P) \right\} \left\{ \prod_{j=1}^k \left[\int \prod_{i=1}^{n_j} P(Y_{ji} \leq y_{ji}) d\Pi(P) \right]\right\}.$$ We overload notation slightly by allowing $\overline\Pi$ to refer to both the probability measure and its CDF. We construct prediction sets $C(\alpha)$ that satisfy $\probnew(Y \in C(\alpha)) \geq 1-\alpha$. Method 0 requires equal $n_j$ across groups, while the other methods allow varying $n_j$. For validity, the non-asymptotic methods (Methods 0, 2, and 3) only require $k\geq 1$ and  $n_j\geq 1$, $j=1,\ldots,k$. On the other hand, some methods place requirements on $k$ and $n_j$ for nontrivial sets. We note these requirements in the theorems associated with each method.

\subsection{Method 0: Double Conformal} \label{section::unsup_double}
The hierarchical set-up involves two levels of randomness. At the level of group $j$, we have independent observations from a distribution $P_j$. At the distribution level, each distribution is sampled from $\Pi$. A ``double conformal'' method is one natural approach that incorporates this hierarchical structure when all $n_j$ values are equal, $j=1,\ldots, k$. (If the samples are not equally sized, then we could work with $\min_j n_j$ observations per group, sampled uniformly at random without replacement.) This method first constructs a prediction set within each group and then uses those sets to construct a final prediction set across groups. 

At the group level, let $C_j(\alpha/2) = [\ell_j, u_j]$ be the $100(1-\alpha/2)\%$ prediction set obtained by applying the method in Theorem~\ref{thm:unsup_basic} at level $\alpha/2$ to group $j$, $j = 1,\ldots, k$. We construct a vector of $k$ lower bounds $(\ell_1, \ldots, \ell_k)$ and $k$ upper bounds $(u_1, \ldots, u_k)$. Using the order statistics from those vectors, we set $C^\double(\alpha) = [\ell_{(r)}, u_{(s)}]$, where $r = \lfloor (k+1)(\alpha/4) \rfloor$ and $s=\lceil (k+1)(1-\alpha/4) \rceil$. If $r < 1$ and $s > k$, let $\ell_{(r)} = \min\{\mathcal{Y}\}$ and $u_{(s)} = \max\{\mathcal{Y}\}$. By Theorem~\ref{thm:unsup_double}, $C^\double(\alpha)$ is a valid $100(1-\alpha)\%$ prediction set for a new $Y$ from a new group.

\begin{restatable}[]{thm}{thmUnsupDouble} \label{thm:unsup_double}
If all groups have an equal number of observations $n_1 = n_2 = \cdots = n_k$, then $\probnew(Y \in C^\double(\alpha))\geq 1-\alpha$ for $C^\double(\alpha)$ as defined above. This method can produce nontrivial sets if $k\geq 4/\alpha - 1$ and $n_1\geq 4/\alpha - 1$.
\end{restatable}

For a proof, see Appendix~\ref{appendix:math}. While this method is valid, our results show that this method overcovers. 
Thus, we turn to several methods that are better choices. 

\subsection{Method 1: Pooling CDFs}

To produce smaller prediction sets, we construct an empirical CDF within each group. We average these CDFs across groups, and we determine the prediction set bounds based on the quantiles of the average of CDFs. If $Y$ has a continuous distribution, this method is asymptotically valid as $k\to\infty$ for any values of $n_j \geq 1$, $j = 1, \ldots, k$.  

Formally, for any group $j$, the empirical CDF is defined as $$\hat{F}_j(t) = \frac{1}{n_j} \sum_{i=1}^{n_j} I(Y_{ji} \leq t).$$ We set 
\begin{align*}
\hat{q}_k(\alpha) &= \inf \left\{ t\in\mathcal{Y} : \frac{1}{k} \sum_{j=1}^k \hat{F}_j(t) \geq \alpha \right\}.
\end{align*}
Then an asymptotic $1-\alpha$ prediction set is $C^\poolCDF(\alpha) = [\hat{q}_k(\alpha/2), \hat{q}_k(1-\alpha/2)].$ For a proof of Theorem~\ref{thm:unsup_pool}, see Appendix~\ref{appendix:math}.
\begin{restatable}[]{thm}{thmUnsupPool} \label{thm:unsup_pool}
Assume that $Y$ has a continuous distribution. For $C^\poolCDF(\alpha)$ as defined above, $\probnew(Y\in C^\poolCDF(\alpha))\to 1-\alpha$ as $k\to \infty$. This method can produce nontrivial sets for any $k\geq 1$ and $n_j\geq 1$, $j=1,\ldots, k$.
\end{restatable}

\subsection{Method 2: Subsampling Once}

While the previous method is asymptotically valid under a CDF condition, we may desire a method without the continuous distribution requirement and with both reasonable coverage and finite sample validity. To achieve those goals, we propose a method based on subsampling. Draw one observation uniformly at random from each group. Then the data consist of $k$ iid observations $Y_1, Y_2, \ldots, Y_k$ from $\tilde{\Pi} = \int P d\Pi(P)$. We define a prediction set $C^\sub(\alpha)=[Y_{(r)}, Y_{(s)}],$ where $r = \lfloor (k+1)(\alpha/2) \rfloor$ and $s = \lceil (k+1)(1-\alpha/2) \rceil$. If $k < 1$ and $s > k$, set $Y_{(r)} = \min\{\mathcal{Y}\}$ and $Y_{(s)} = \max\{\mathcal{Y}\}$. Since the subsample contains $k$ iid draws from $\tilde{\Pi}$, Theorem~\ref{thm:unsup_once} follows from Theorem~\ref{thm:unsup_basic}.

\begin{restatable}[]{thm}{thmUnsupOnce} \label{thm:unsup_once}
For $C^\sub(\alpha)$ as defined above, $\probnew(Y \in C^\sub(\alpha))\geq 1-\alpha$. This method can produce nontrivial sets if $k\geq 2/\alpha - 1$ and $n_j \geq 1$, $j=1,\ldots,k$.
\end{restatable}

\subsection{Method 3: Repeated Subsampling} \label{section::unsup_repeated}

The single subsampling approach is simple and valid, but it ignores most of the data. Due to the use of a single subsample, the results may be insufficiently reproducible. We address these shortcomings by incorporating $B$ subsamples of a single observation from each of the $k$ groups. \cite{gupta2020nested} developed the method of constructing conformal prediction sets through repeated subsampling in the case of exchangeable data. Suppose $Y^b_{(1)}, Y^b_{(2)}, \ldots, Y^b_{(k)}$ are the ordered observations from the $b^{th}$ subsample. Conformal prediction is implicitly testing $H_0: Y_{k+1} = u$ versus $H_1: Y_{k+1}\neq u$, and the level $1-\alpha$ conformal prediction set is the set of values at which we would not reject $H_0$ under the given construction. Thus, within the $b^{th}$ subsample, the $p$-value at $u\in\mathcal{Y}$ is
\[\pi_b(u) = \begin{cases} 
      1 & \text{if } u \in [Y^b_{(\lfloor (k+1)/2 \rfloor)}, Y^b_{(\lceil (k+1)/2 \rceil)}] \\
      \inf\{\alpha : u\notin [Y_{(r)}^b, Y_{(s)}^b] \} & \text{otherwise}
   \end{cases},
\]
where $r = \lfloor (k+1)(\alpha/2) \rfloor$ and $s = \lceil (k+1)(1-\alpha/2) \rceil$. We define a prediction set $C^\rep(\alpha) = \left\{u: B^{-1}\sum_{b=1}^B \pi_b(u) \geq \alpha\right\}.$
\begin{restatable}[]{thm}{thmUnsupRep} \label{thm:unsup_rep}
For $C^\rep(\alpha)$ as defined above, $\probnew(Y \in C^\rep(\alpha)) \geq 1-2\alpha$. This method can produce nontrivial sets if $k > 2/\alpha - 1$ and $n_j\geq 1$, $j=1,\ldots,k$.
\end{restatable}

Theorem~\ref{thm:unsup_rep} holds because $(2/B) \sum_{b=1}^B \pi_b(u)$ (double the test statistic) is a valid $p$-value for the stated test \citep{ruschendorf1982random, meng1994posterior, barber2021predictive, vovk2020}. In practice, however,  $C^\rep(\alpha)$ has close to $100(1-\alpha)\%$ coverage. The guaranteed level $1-2\alpha$ coverage and empirical level $1-\alpha$ coverage is analogous to the coverage of the jackknife+ method \citep{barber2021predictive}, which constructs conformal sets through leave-one-out prediction. Furthermore, \cite{tian2021large} show that the multiplicative correction is not always necessary when averaging $p$-values. Their Corollary~1 proves  that for small enough $\alpha$, the average of $p$-values on one-dimensional normal random variables with arbitrary positive correlation (and certain degrees of negative correlation) is a valid $p$-value. For a proof of the nontrivial condition, see Appendix~\ref{appendix:math}.

\subsection{Unsupervised New Distribution Simulations} 
\label{section::simulations_unsup}

To understand the performance of these methods, we consider a simulation study. We begin by generating data from $k$
distributions. We draw $\theta_1, \ldots, \theta_k\sim N(0,1)$. Then we simulate $Y_{j1}, \ldots,
Y_{jn_j}\sim N(\theta_j, 1)$ for $j = 1, \ldots, k$. 
We use $n_j = 100$ observations per group. We vary the number of groups ($k$) from 5 to 100 in increments of 5 and from 200 to 1000 in increments of 100.  The repeated subsampling sets use $B=100$ subsamples. Each simulation generates a data sample, draws a new $\theta_{k+1}\sim N(0,1)$ and  $Y \sim N(\theta_{k+1}, 1)$, constructs a prediction set $C(\alpha)$, determines the size of the prediction set, and checks whether $Y\in C(\alpha)$. The coverage is the proportion of simulations for which $Y\in C(\alpha)$. We set $\alpha=0.1$,
and we perform 1000 simulations at each value of $k$. 

\begin{figure}[htp]
\centering
\begin{subfigure}{\textwidth}
\includegraphics[scale=.6]{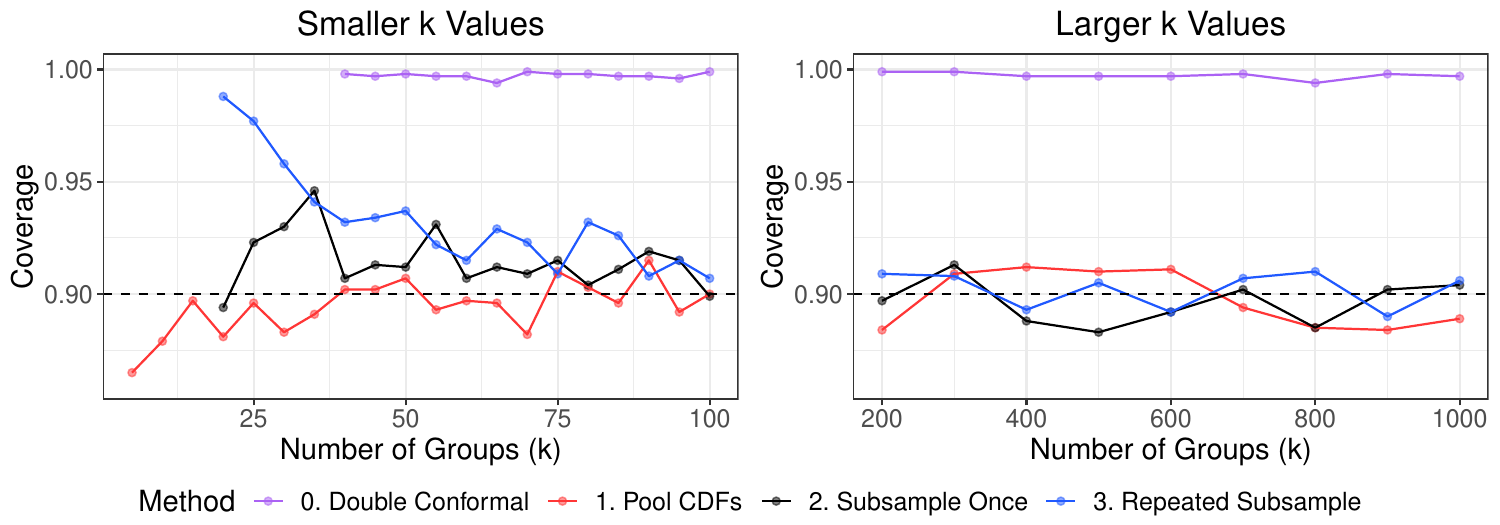}
\subcaption{Coverage of unsupervised conformal prediction sets for an outcome from a new group.}
\label{fig:unsup_cov}
\end{subfigure}
\begin{subfigure}{\textwidth}
\includegraphics[scale=.6]{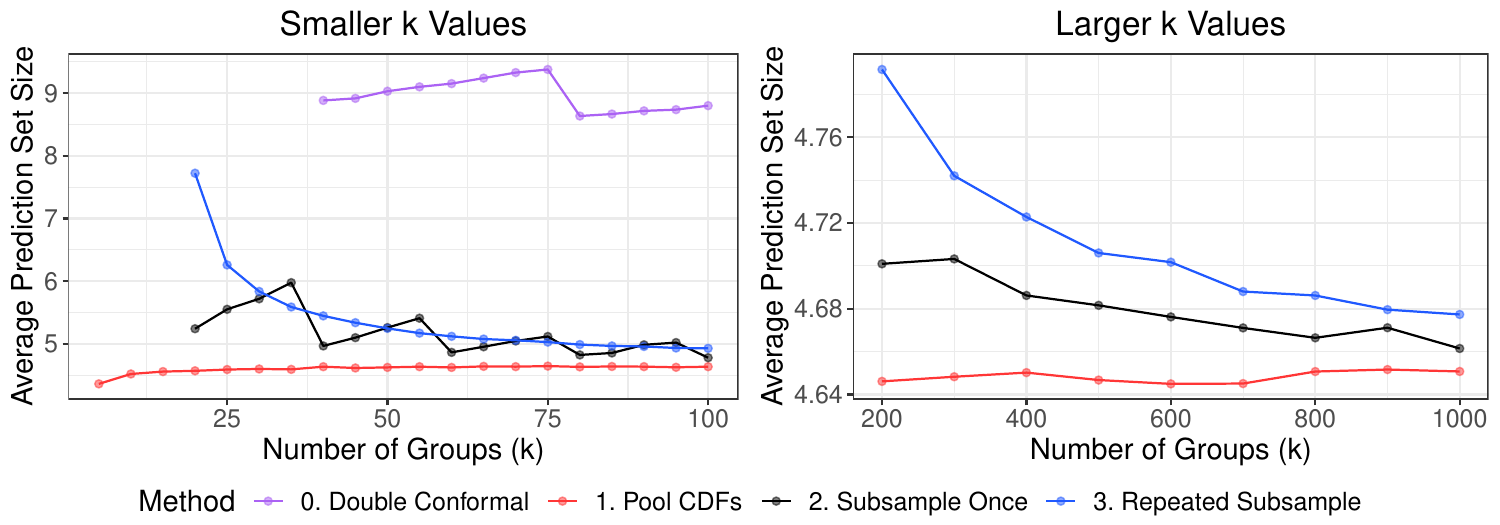}
\subcaption{Average size of unsupervised conformal prediction sets for an outcome from a new group. }
\label{fig:unsup_size}
\end{subfigure}
\caption{Unsupervised conformal prediction simulations in a setting with balanced groups. CDF pooling produces the smallest prediction sets. This method is asymptotically valid as $k\to\infty$ and has approximately nominal coverage in simulations.}
\label{fig:unsup_coverage_size}
\end{figure}

Figure \ref{fig:unsup_coverage_size} displays the empirical coverage and average set length from the four unsupervised methods.  The double conformal method consistently overcovers, with coverage close to 1. CDF pooling undercovers at small to moderate values of $k$ (e.g., $k\leq 35$) but has approximately $1-\alpha$ coverage for larger $k$. Single and repeated subsampling tend to overcover for small to moderate $k$ and have approximately $1-\alpha$ coverage for large $k$.

The pooling method has the smallest sets, the single subsampling and repeated subsampling methods have the next largest sets (mostly on par), and double conformal has the largest sets. (The right panel of Figure~\ref{fig:unsup_size} excludes the double conformal sets, which have average lengths between 8.4 and 8.6 for $k\geq 200$.) Appendix~\ref{appendix:sims} contains simulations that produce similar behavior on normal data at $n_j\in\{40, 1000\}$ and on non-normal data.

\begin{figure}[htp]
\centering
\includegraphics[scale=.6]{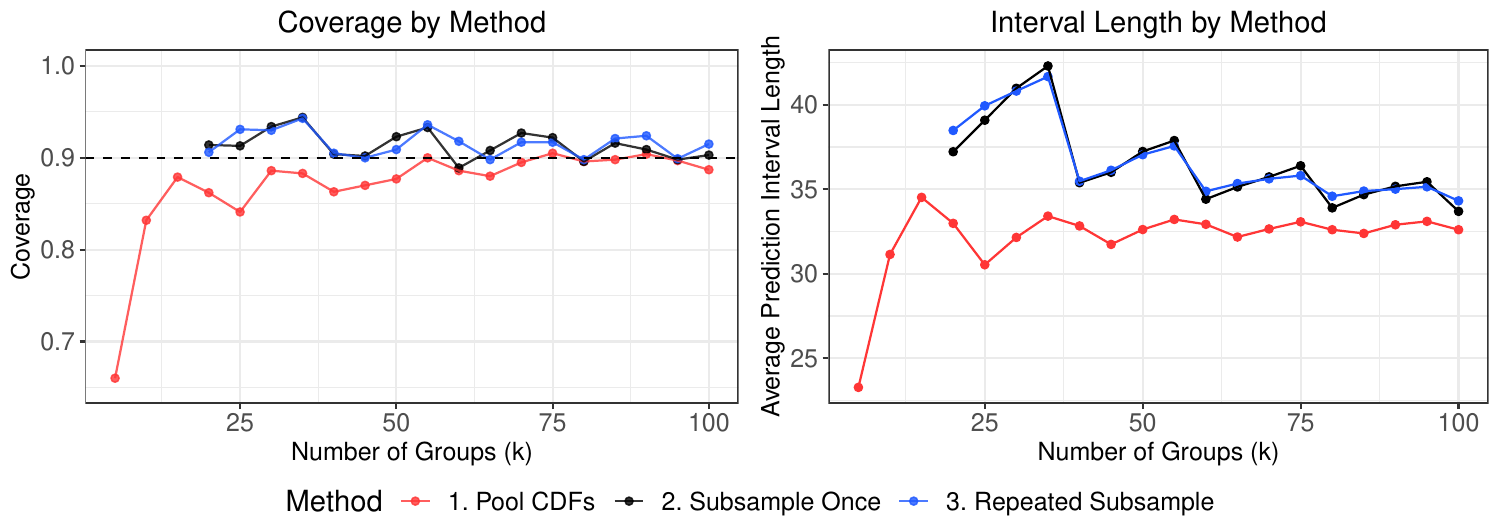}
\caption{In this unbalanced unsupervised setting, one group has 1000 observations, and the remaining $k-1$ groups have 5 observations. CDF pooling undercovers for small $k$.}
\label{fig:unbalanced}
\end{figure}

Figure~\ref{fig:unsup_coverage_size} has considered cases with balanced numbers of observations in each group. We now consider a highly unbalanced case: one group has 200 times as many observations as each of the other groups, and the between-group variation exceeds the within-group variation by three orders of magnitude.
We take
$Y_{ij}\sim N(\theta_j, \sigma^2 = 0.1)$
where $\theta_j \sim N(0, \tau^2 = 100)$,
$n_1 = 1000$,
and $n_j =5$ for $2\leq j \leq k$.
We let $k$ vary from 5 to 100 in increments of 5. 
Figure \ref{fig:unbalanced} shows that the single subsample and repeated subsample methods typically have at least nominal coverage. CDF pooling undercovers for small $k$. For $k\geq 20$, all three methods produce finite prediction intervals, and CDF pooling only undercovers by about 0.05. CDF pooling produces the smallest prediction sets, and single and repeated subsampling have similar average prediction set lengths. Thus, the behavior we observe in this highly unbalanced case is similar to the balanced case.

Overall, CDF pooling appears to be the best choice, with a few caveats. CDF pooling consistently produces the smallest prediction sets, and it achieves nominal or approximately nominal coverage. As a drawback, this method often slightly undercovers for small to moderate $k$, with more notable undercoverage in the small $k$ unbalanced setting. In addition, this method only guarantees coverage asymptotically as $k\to\infty$ and for continuous $Y$. Hence, if we desire a method with consistent simulated coverage, with theoretical guarantees on coverage, or without the CDF requirement, the subsampling conformal methods are better choices. Between the two subsampling methods, we recommend repeated subsampling. This method has guaranteed coverage at level $1-2\alpha$, but in practice it tends to cover at level $1-\alpha$. Furthermore, for moderate $k$, its prediction sets are about the same size as the single subsample method. Favorably, repeated subsampling yields more reproducible prediction intervals than single subsampling. To understand the variation in these two methods, we consider a single dataset with $k=100$ groups and $n_j = 100$ observations per group, using the same setup as Figure~\ref{fig:unsup_coverage_size}. Based on Figure~\ref{fig:unsup_coverage_size}, these methods have similar coverage and size at these parameters. Through 1000 repetitions, we construct $90\%$ prediction intervals using the two subsampling methods. Across simulations, single subsampling has lower bounds between $-3.6$ and $-1.7$, while repeated subsampling has lower bounds between $-2.7$ and $-2.4$. Similarly, single subsampling has upper bounds between $1.5$ and $3.5$, while repeated subsampling has upper bounds between $2.4$ and $2.7$. 
Thus, as expected, we see less variation in the prediction intervals constructed through repeated subsampling.

\section{SUPERVISED PREDICTION FOR A NEW DISTRIBUTION}
\label{section::sup_methods}

In the supervised case, each group $\mathcal{D}_1, \ldots, \mathcal{D}_k$ has iid data $\mathcal{D}_j = \{(X_{j1}, Y_{j1}), \ldots, (X_{jn_j}, Y_{jn_j}) \} \sim P_j$, where $Y_{j1}, \ldots, Y_{jn_j} \in \mathcal{Y} \subseteq \real$ and $P_1,\ldots, P_k\sim\Pi$. Each $X_{ji}$ is a $p$-dimensional vector given by $X_{ji} = (X_{ji}^{(1)}, X_{ji}^{(2)}, \ldots, X_{ji}^{(p)})$. Suppose we have a new distribution $P_{k+1}\sim\Pi$ and $(X,Y)\sim P_{k+1}$. Then $(X, Y) \sim \tilde{\Pi}$, where $\tilde{\Pi} = \int P d\Pi(P)$. Assuming that we only observe $X=x$, we want a prediction region for $Y=y$. To define a distribution, we use $x \in \real^p$, $y\in\mathcal{Y}$, $\mathbf{x}_j \in \real^{n_j \times p}$, and $y_j \in \mathcal{Y}^{n_j}$. Similar to the unsupervised setting, we define a distribution over the randomness in the initial sample and the new $(X,Y) \sim \tilde{\Pi}$ as 
\begin{align*}
\overline{\Pi}&(x, y, \mathbf{x}_1, y_1, \mathbf{x}_2, y_2, \ldots, \mathbf{x}_k, y_k) = \left\{ \int P(X^{(1)} \leq x^{(1)}, \ldots, X^{(p)} \leq x^{(p)}, Y \leq y) d\Pi(P) \right\} \times \\
&\quad \left\{ \prod_{j=1}^k \left[\int \prod_{i=1}^{n_j} P(X_{ji}^{(1)} \leq \mathbf{x}_{ji}^{(1)}, \ldots, X_{ji}^{(p)} \leq \mathbf{x}_{ji}^{(p)}, Y_{ji} \leq y_{ji}) d\Pi(P) \right]\right\}.
\end{align*}
We construct sets $C(x;\alpha)$ such that $\probnew(Y \in C(X;\alpha)) \geq 1-\alpha$ for $(X, Y)\sim \tilde{\Pi}$. As in the unsupervised case, the non-asymptotic, subsampling methods are valid for $k\geq 1$ and $n_j \geq 1$, $j = 1,\ldots, k$. We note requirements on $k$ and $n_j$ for nontrivial sets. 

\subsection{Method 1: Pooling CDFs}
Similar to the unsupervised setting, we consider methods that average empirical CDFs across groups. We first consider a sample-splitting method that is asymptotically valid as $k\to\infty$, regardless of the choice of model. Let $[k] = \{1, \ldots, k\}$. We start by using the observations from some strict subset $k_0 \subset [k]$ of the $k$ groups to fit any model $\hat{\mu}(X)$ as an estimator of $\E[Y \mid X]$. For instance, $\hat{\mu}(X)$ could be a single model based on the pooled observations or an average of $k_0$ models fit on the individual groups. Importantly, $\hat{\mu}(X)$ must stay fixed as $k$ grows. Since $\hat{\mu}(x)$ will be the center of the prediction set, it is best if $\hat{\mu}(X)$ is a good approximation to $\E[Y \mid X]$. We use the remaining groups to fit the residuals $R_{ji} = |Y_{ji} - \hat{\mu}(X_{ji})|$, $j\in [k]\backslash k_0$, $i = 1,\ldots, n_j$. Now for each $j\in [k]\backslash k_0$, we define group $j$'s empirical CDF of the residuals $$\hat{F}_j(t) = \frac{1}{n_j} \sum_{i=1}^{n_j} I(R_{ji} \leq t).$$ We define $$\hat{q}_k(\alpha) = \inf\left\{t\in\real: \frac{1}{|[k]\backslash k_0|} \sum_{j\in [k]\backslash k_0} \hat{F}_j(t) \geq \alpha \right\}.$$  For continuous $Y$, $C^\poolCDF(x; \alpha) = [\hat{\mu}(x) - \hat{q}_k(1-\alpha), \hat{\mu}(x) +  \hat{q}_k(1-\alpha)]$ is an asymptotic $1-\alpha$ prediction set. For a proof of Theorem~\ref{thm:sup_pool}, see Appendix~\ref{appendix:math}.
\begin{restatable}[]{thm}{thmSupPool} \label{thm:sup_pool} 
Fit a model $\hat{\mu}(X)$ as an estimator of $\E[Y \mid X]$ using the observations in groups \mbox{$k_0 \subset [k]$.} ($\hat{\mu}(X)$ stays fixed as $k$ grows.) If $Y$ has a continuous distribution, then $\probnew(Y \in C^\poolCDF(X; \alpha)) \overset{p}{\to} 1-\alpha$ as $k\to\infty$. At any $x$, this method can produce nontrivial sets for $k\geq 1$ and $n_j\geq 1$, $j=1,\ldots, k$.
\end{restatable}

Under stronger assumptions on the agreement between the true and estimated models, we consider a second asymptotically valid approach ($k\to\infty$) that does not require sample splitting. If $(X_j, Y_j)\sim P_j$, then suppose $Y_j = \mu_{P_j}(X_j) + \epsilon$, where $\epsilon$ has a zero-mean distribution. For each group $j\in\{1,\ldots,k\}$, we use the observations $X_{j1}, X_{j2}, \ldots X_{jn_j}$ to fit a model $\hat{\mu}_{P_j}$. At any given $x$, we define a pooled model $\mu(x) = \int \mu_P(x) d\Pi(P)$ and an estimated pooled model $\hat{\mu}(x) = k^{-1} \sum_{j=1}^k \hat{\mu}_{P_j}(x).$
Under $\mu$ and $\hat{\mu}$, we have the residuals $R_{ji}(\mu) = \left|\mu(X_{ji}) - Y_{ji} \right|$ and $R_{ji}(\hat{\mu}) = \left|\hat{\mu}(X_{ji}) - Y_{ji} \right|.$
The residuals have empirical CDFs
\begin{align*}
\hat{F}_{j,\mu}(t) &= \frac{1}{n_j} \sum_{i=1}^{n_j} I(R_{ji}(\mu) \leq t) \\
\hat{F}_{j,\hat{\mu}}(t) &= \frac{1}{n_j} \sum_{i=1}^{n_j} I(R_{ji}(\hat{\mu}) \leq t).
\end{align*}  
We obtain sample quantiles
$$\hat{q}_k(\hat{\mu}; \alpha) = \inf\left\{t\in\mathbb{R}: \frac{1}{k}\sum_{j=1}^k \hat{F}_{j,\hat{\mu}}(t) \geq \alpha \right\}.$$
Under the assumptions on $\mu$, $\hat{\mu}$, and $\tilde{\Pi}$ stated in Theorem~\ref{thm:sup_model_avg}, an asymptotic $1-\alpha$ prediction set is $C^\modelAvg(x; \alpha) = [\hat{\mu}(x) - \hat{q}_k(\hat{\mu}; 1-\alpha), \hat{\mu}(x) + \hat{q}_k(\hat{\mu}; 1-\alpha)]$.

\begin{restatable}[]{thm}{thmSupModelAvg} \label{thm:sup_model_avg}
Suppose $Y\sim\tilde{\Pi}$ has a continuous distribution.  If $(X_j, Y_j) \sim P_j$, suppose  $Y_j = \mu_{P_j}(X_j) + \epsilon$, where $\epsilon$ has a zero-mean distribution. Assume $\hat{\mu}$ satisfies
$$\frac{1}{k}\sum_{j=1}^k \sup_t |\hat F_{j,\hat\mu}(t)-\hat F_{j,\mu}(t)| \overset{p}{\to} 0 $$
as $k\to\infty$, and assume that for $\delta > 0$, $\lim_{k\to\infty} \probnew(|\hat{\mu}(X) - \mu(X)| > \delta) = 0$. For $C^\modelAvg(x; \alpha)$ as defined above, $\lim_{k\to\infty} \probnew(Y \in C^\modelAvg(X; \alpha)) = 1-\alpha$. At any value of $x$, this method can produce nontrivial sets for $k\geq 1$ and $n_j\geq 1$, $j=1,\ldots, k$.
\end{restatable}

\subsection{Method 2: Subsampling Once} \label{section::sup_once}
If we desire a method with finite sample coverage guarantees, a conformal method based on subsampling can achieve that goal. As in the unsupervised setting, we randomly select one observation from each of the $k$ groups. This creates a sample of $k$ pairs of iid observations $(X,Y)$. Suppose we have a new data point $(X_{k+1}, Y_{k+1})\sim P_{k+1}$, but we only observe $X_{k+1}$.
Letting $X_{k+1} = x$,
we have an augmented $X$ sample $(X_1,\ldots,X_k, X_{k+1})$. For each possible $y$, we test $H_0: Y_{k+1} = y$ 
at a $1-\alpha$ confidence
level using the following procedure: Assume
$Y_{k+1} = y$, giving an augmented $Y$ sample of $(Y_1, \ldots, Y_k,
Y_{k+1})$. Using the sample augmented with $(x,y)$ as training data, fit a model
$\hat{\mu}_{(x,y)}(X)$ as an estimator of $\E[Y \mid X]$.  
Then compute nonconformity
scores $R_i(x,y) = |\hat{\mu}_{(x,y)}(X_i) - Y_i|$, $i=1,\ldots, k+1$. The
$p$-value for the test of $H_0: Y_{k+1} = y$ is $\pi(x,y) =
(k+1)^{-1}\sum_{i=1}^{k+1} I(R_i(x,y) \geq R_{k+1}(x,y))$. The $1-\alpha$ conformal prediction
set is $C^\sub(x; \alpha) = \{y\in\mathbb{R}: \pi(x,y)\geq \alpha\}$. 
\begin{restatable}[]{thm}{thmSupOnce} \label{thm:sup_once}
For $C^\sub(x; \alpha)$ as defined above, $\probnew(Y \in C^\sub(X; \alpha)) \geq 1-\alpha$.  For $C^\sub(x; \alpha)$ to be nontrivial, it must hold that $k > 1/\alpha - 1$ and $n_j \geq 1$, $j = 1, \ldots, k$.
\end{restatable}
Since the subsample of $k$ observations is an iid sample, Section~\ref{section::basic} justifies this method. Note that for any $(x,y)$, $\pi(x,y) \geq 1/(k+1)$. Hence, if $k \leq 1/\alpha - 1$, then $\pi(x,y) \geq 1/(k+1) \geq \alpha$ will hold for all $(x,y)$. 
% The nontrivial condition is necessary but not sufficient.

\subsection{Method 3: Repeated Subsampling}

Since a single subsample ignores most of the data, it may be inadequately reproducible. To improve the reproducibility, we modify the previous method to incorporate $B$ subsamples of a single observation from each of the $k$ groups. For the $b^{th}$ subsample, $(X_{1b}, Y_{1b}),  \ldots, (X_{kb}, Y_{kb})$ contains one observed pair from each of the $k$ groups. Suppose $X_{k+1} = x$ is a new observation from a new group. Conformal prediction is implicitly testing $H_0: Y_{k+1} = y$ versus $H_1: Y_{k+1}\neq y$, and the level $1-\alpha$ conformal prediction set is the set of values at which we would not reject $H_0$ under the given construction. Using the $b^{th}$ subsample augmented with $(x,y)$, we construct residuals $R_{b,i}(x,y)$ in the same manner as Section~\ref{section::sup_once}. Then $\pi_b(x,y) = (k+1)^{-1} \sum_{i=1}^{k+1} I(R_{b,i}(x,y) \geq R_{b,k+1}(x,y))$ is a valid $p$-value for the stated test. We construct $\pi_b(x,y)$ for $B$ subsamples. We define $C^\rep(x; \alpha) = \left\{y: B^{-1}\sum_{b=1}^B \pi_b(x,y) \geq \alpha\right\}.$
\begin{thm} \label{thm:sup_rep}
For $C^\rep(x; \alpha)$ as defined above, $\probnew(Y \in C^\rep(X; \alpha)) \geq 1-2\alpha$. For $C^\rep(x; \alpha)$ to be nontrivial, it must hold that $k > 1/\alpha - 1$ and $n_j \geq 1$, $j = 1, \ldots, k$.
\end{thm}
Similar to Theorem~\ref{thm:unsup_rep} in the unsupervised case, Theorem~\ref{thm:sup_rep} is true because $(2/B) \sum_{b=1}^B \pi_b(x, y)$ is a valid $p$-value for the stated test. As in the unsupervised case, $C^\rep(x; \alpha)$ has empirical coverage of approximately $1-\alpha$. The nontrivial condition in Theorem~\ref{thm:sup_rep} holds for the same reason as the nontrivial condition in Theorem~\ref{thm:sup_once}.

\subsection{Supervised New Distribution Simulations}
\label{section::simulations_sup}

We explore the supervised prediction methods through a simulation study.
To generate data from $k$ distributions, we draw
$\theta_1,\ldots,\theta_k \sim N(\mu,\tau^2)$,
$X_{j1},\ldots,X_{jn_j} \sim N(0,1)$, and
$\epsilon_{j1},\ldots,\epsilon_{jn_j} \sim N(0,1)$.
We let $Y_{ji} = \theta_j X_{ji} + \epsilon_{ji}$, $j=1,\ldots,k$, $i = 1,\ldots,n_j$. 
Then we draw a new $X\sim N(0,1)$,
$\theta_{k+1}\sim N(\mu,\tau^2)$, and response $Y \sim N(\theta_{k+1} X, 1)$. 
Treating $\theta_{k+1}$ and $Y$ as
unknown, we wish to predict $Y$ from the observed $X = x$.
For the CDF pooling method, we use the approach justified by Theorem~\ref{thm:sup_pool}. We pool the observations from $k_0 = \lfloor k/2 \rfloor$ groups to fit a one-parameter linear regression model $\hat{\mu}(X) = \hat\theta X$. Then we use the remaining groups for quantile estimation.
For the subsampling methods, we fit $\hat{\mu}_{(x,y)}(X;\hat{\theta}) = \hat\theta X$ using subsamples of one observation per group, augmented with $(x,y)$. For repeated subsampling, we use $B = 100$ subsamples.

We draw $n_j = 100$ observations per group. We vary the number of groups ($k$) from 20 to 100 in increments of 5 and from 200 to 1000 in increments of 100. To draw the
$\theta$ parameters, we set $\mu = 0$ and $\tau^2 = 1$. In Appendix~\ref{appendix:sims}, we see similar behavior for $n_j \in \{20, 1000\}$ and for $\mu = 1$ and $\tau^2 = 0.1$. We perform 1000 simulations at each $k$. We set $\alpha = 0.1$. Each simulation generates a data sample, draws a new $(X,Y)$ from a new distribution, constructs a prediction set $C(X; \alpha)$, determines the size of the set, and checks whether $Y\in C(X; \alpha)$.

\begin{figure}[htp]
\centering
\begin{subfigure}{\textwidth}
\includegraphics[scale=.6]{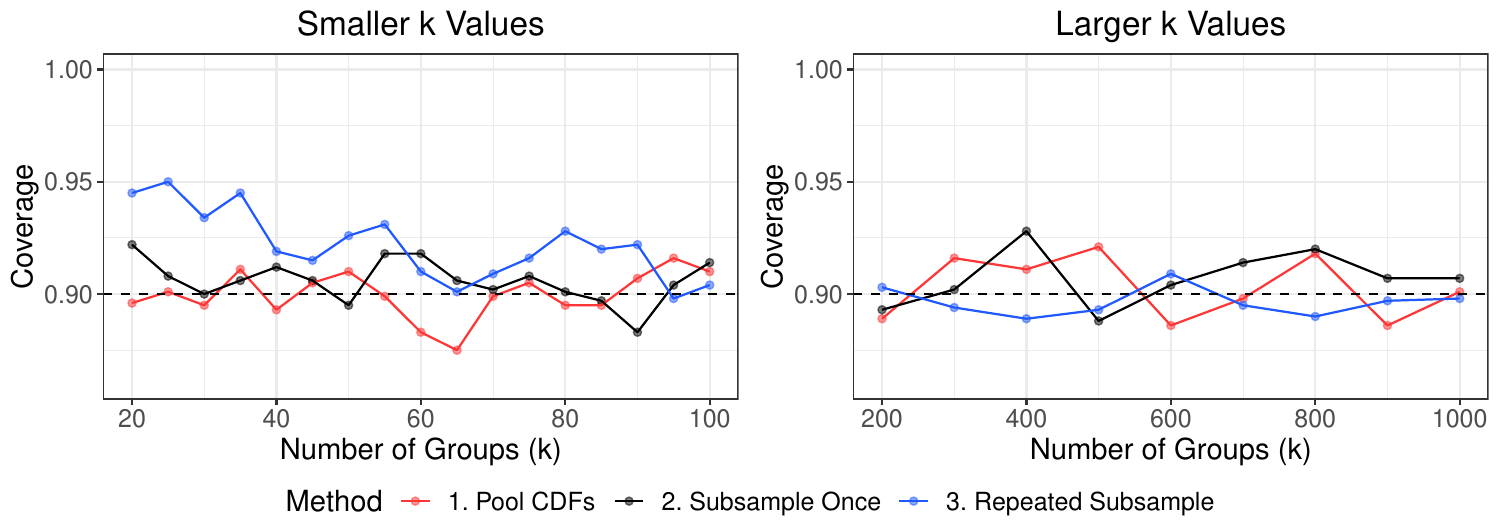}
\subcaption{Coverage of supervised conformal prediction sets for an outcome from a new group.}
\end{subfigure}
\begin{subfigure}{\textwidth}
\includegraphics[scale=.6]{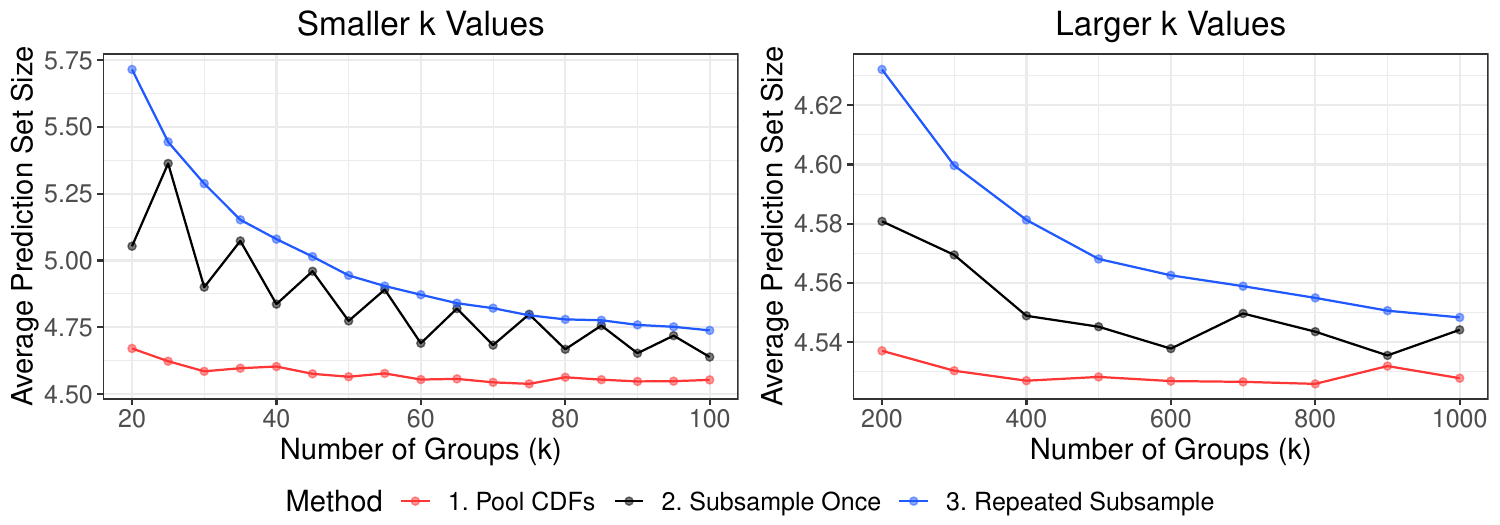}
\subcaption{Average size of supervised conformal prediction sets for an outcome from a new group. }
\end{subfigure}
\caption{All supervised conformal prediction methods have approximately nominal coverage in simulations. CDF pooling sets are the smallest, followed by single subsampling and repeated subsampling.}
\label{fig:sup_coverage_size}
\end{figure}

Figure \ref{fig:sup_coverage_size} shows the coverage and average length of the prediction sets from these supervised methods. The coverage is the proportion of simulations for which $Y\in C(X; \alpha)$. All three methods have coverage close to $1-\alpha$ for all $k$. For small $k$, repeated subsampling often overcovers by up to 0.05. The pooling sets are the smallest, followed by the single subsampling sets, and the repeated subsampling sets are the largest. CDF pooling appears to be the best choice in this setting, with the caveats that it has asymptotic coverage and assumes $Y$ is continuous. If choosing between the subsampling methods, we recommend repeated subsampling. While single subsampling often produces slightly smaller prediction sets, the average size differs by less than $1\%$ for $k\geq 300$ in our simulations. Furthermore, the results from repeated subsampling are more reproducible. Similar to the unsupervised case, we use the setup from Figure~\ref{fig:sup_coverage_size} to create a single dataset with $k=100$ groups and $n_j = 100$ observations per group. In 1000 repetitions, we construct $90\%$ prediction intervals at $X=1$ using the two subsampling methods. Across simulations, single subsampling has lower bounds between $-3.7$ and $-1.2$, while repeated subsampling has lower bounds between $-2.7$ and $-2.4$. Similarly, single subsampling has upper bounds between $1.6$ and $3.6$, while repeated subsampling has upper bounds between $2.4$ and $2.7$. Again, the repeated subsampling intervals have less variation than the single subsampling intervals.

\section{UNSUPERVISED PREDICTION FOR AN OBSERVED GROUP} \label{section::obs_methods}

Task 2 considers unsupervised prediction of a
new observation on an existing subject. For each subject $j$, we observe $n_j$ iid real-valued samples $Y_{j1}, \ldots, Y_{jn_j} \sim P_j$. We assume $P_1, \ldots, P_k \sim \Pi$.   We assume without loss of generality that we wish to predict a new observation from subject 1. Hence, this setting's probability distribution $\overline{\Pi}^{(1)}$ accounts for the randomness in both the training sample and this new observation $Y_{1, n_1 + 1}\sim P_1$. For $y_1 \in \real^{n_1 + 1}$ and $y_j \in \real^{n_j}, 2\leq j\leq k$, we define
$$\overline{\Pi}^{(1)}(y_1, y_2, \ldots, y_k) = \left\{ \int \prod_{i=1}^{n_1 + 1} P(Y_{1i} \leq y_{1i}) d\Pi(P) \right\} \left\{ \prod_{j=2}^k \left[\int \prod_{i=1}^{n_j} P(Y_{ji} \leq y_{ji}) d\Pi(P) \right]\right\}.$$ Our conformal prediction sets $C(\alpha)$ satisfy $\probnewone(Y_{1, n_1 + 1} \in C(\alpha)) \geq 1-\alpha$.

We explore two conformal methods to capture the new observation from subject 1. The first method is a standard conformal procedure using subject 1's data. The second method ``borrows'' information from other subjects to obtain a shrinkage estimator of the mean of subject 1's data. Then it performs conformal prediction using this shrinkage estimator. The validity of either method follows
from the usual theory described in Section~\ref{section::basic},
but the shrinkage approach may lead to
smaller prediction sets. While validity holds for $n_1 \geq 1$, both methods require $n_1 > 1/\alpha - 1$ for nontrivial sets.

\subsection{Method 1: Isolate Single Group}

The simplest approach constructs conformal prediction sets for subject 1 using only subject 1's observed data. We propose a new $y$, and we wish to test $H_0: Y_{1,n_1+1} = y$ at a $1-\alpha$ confidence level. Letting $Y_{1,n_1+1} = y$, we have an augmented data vector $(Y_{1,1}, \ldots, Y_{1,n_1}, Y_{1,n_1+1})$ for subject 1. We define $\overline{Y}_{1} = (1/(n_1 + 1)) \sum_{i=1}^{n_1+1} Y_{1,i}$. Then we calculate nonconformity scores $R_i = \left|Y_{1,i} - \overline{Y}_{1}\right|$, $i=1,\ldots,n_1+1$. The $p$-value for the test of $H_0: Y_{1,n_1+1} = y$ is $\pi(y) = (1/(n_1 + 1)) \sum_{i=1}^{n_1+1} I(R_i \geq R_{n_1+1})$. We invert this test to obtain a $1-\alpha$ conformal prediction set $C^\isolate(\alpha) = \{y: \pi(y)\geq\alpha\}$. Since this approach uses conformal methods on iid observations from a single distribution, Theorem~\ref{thm::basic} justifies Theorem~\ref{thm:obs_isolate}.

\begin{restatable}[]{thm}{thmObsIsolate} \label{thm:obs_isolate}
For $C^\isolate(\alpha)$ as defined above, $\probnewone(Y_{1,n_1+1} \in C^\isolate(\alpha)) \geq 1-\alpha$. For $C^\isolate(\alpha)$ to be nontrivial, it must hold that $n_1 > 1/\alpha - 1$.
\end{restatable}

\subsection{Method 2: James-Stein Shrinkage}

A conformal method that borrows strength across distributions may yield tighter prediction intervals, especially if the $k$ distributions are ``close'' to each other. To use the data from all subjects, we work with a conformal residual based on shrinkage. Again, we propose a new value of $y$, and we wish to test $H_0: Y_{1,n_1+1} = y$ at a $1-\alpha$ confidence level. We define $\overline{Y}_{1} = (1 / (n_1 + 1)) \sum_{i=1}^{n_1+1} Y_{1,i}$. Then for $j=2,\ldots,k$, we define $\overline{Y}_{j} = n_j^{-1} \sum_{i=1}^{n_j} Y_{j,i}$. Let $\nu = k^{-1} \sum_{j=1}^k \overline{Y}_j$, and let $\widehat{\sigma}_1^2$ be the sample variance of $(Y_{1,1}, \ldots, Y_{1,n_1}, Y_{1,n_1+1})$.

Now in place of $\overline{Y}_1$ in the nonconformity scores, we use the James-Stein shrinkage estimator: $$\widetilde{Y}_1 = \left(1 - \frac{(k-2)\widehat{\sigma}_1^2/(n_1+1)}{\sum_j (\overline{Y}_j - \nu)^2} \right)_+ (\overline{Y}_1 - \nu) + \nu,$$
where $(x)_+ = \max(x, 0).$  $\tilde{Y}_1$ is defined when $k\geq 2$. The rest of the procedure mirrors the previous method. We calculate nonconformity scores $R_i = |Y_{1,i} - \widetilde{Y}_1|$, $i=1,\ldots, n_1+1$. For the proposed $y$, we obtain a $p$-value $\pi(y) = (1 / (n_1+1)) \sum_{i=1}^{n_1+1} I(R_i \geq R_{n_1+1})$. The $1-\alpha$ conformal prediction set is $C^\shrinkage(\alpha) = \{y: \pi(y)\geq\alpha\}$. Theorem~\ref{thm::basic} justifies Theorem~\ref{thm:obs_shrinkage}.

\begin{restatable}[]{thm}{thmObsShrinkage} \label{thm:obs_shrinkage}
For $C^\shrinkage(\alpha)$ as defined above, $\probnewone(Y_{1,n_1+1} \in C^\shrinkage(\alpha)) \geq 1-\alpha$. For $C^\shrinkage(\alpha)$ to be nontrivial, it must hold that $n_1 > 1/\alpha - 1$ and $k\geq 2$.
\end{restatable}

\subsection{Unsupervised Observed Group Simulations}
\label{section::borrow}

We compare these methods under two data generation processes. We draw subject-specific means $\theta_1,\ldots,\theta_k\sim N(0,1)$. Then for $j=1,\ldots,k$, we generate $Y_{j1}, Y_{j2},\ldots, Y_{jn_j} \sim N(\theta_j, \sigma^2)$. We consider $\sigma^2 = 1$ and $\sigma^2 = 100$. Across all
simulations, we set $n_j=20$. We vary $k$ from 5 to 1000 in increments
of 5. At each choice of $k$, we perform 1000 simulations at $\alpha = 0.1$. Each simulation generates a data sample, draws another observation $Y_{1,n_1+1} \sim N(\theta_1, \sigma^2)$ from subject 1's distribution, constructs a prediction set $C(\alpha)$ for subject 1, determines the size of the prediction set, and checks whether $Y_{1,n_1+1} \in C(\alpha)$.

\begin{figure}[htp]
\centering
\includegraphics[scale=0.6]{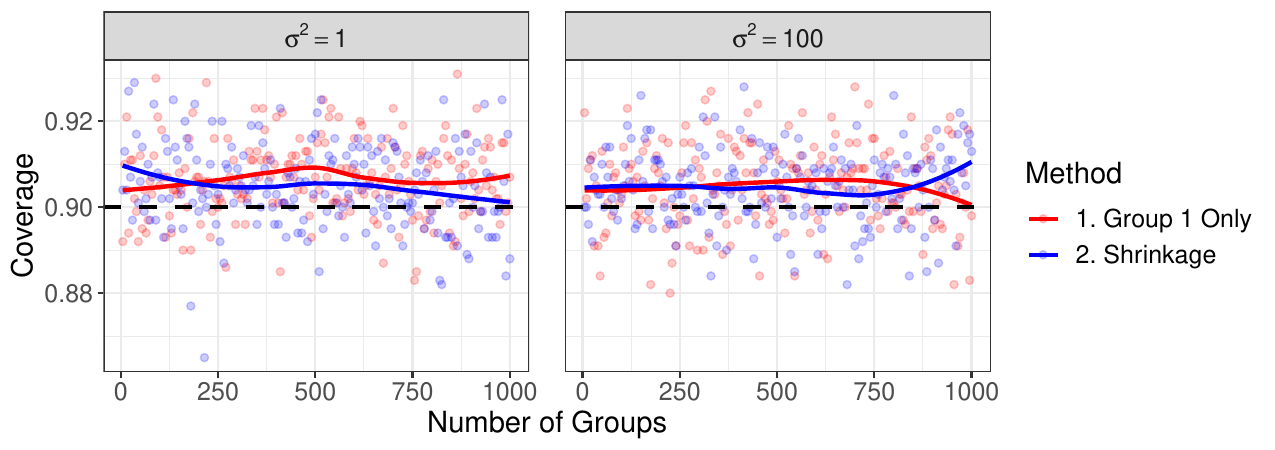}
\caption{Coverage of conformal methods for a new observation from an observed group. Loess smoothing for visualization. Both methods have approximately nominal coverage.}
\label{fig:shrinkage_cov}
\end{figure}

\begin{figure}[htp]
\centering
\includegraphics[scale=0.6]{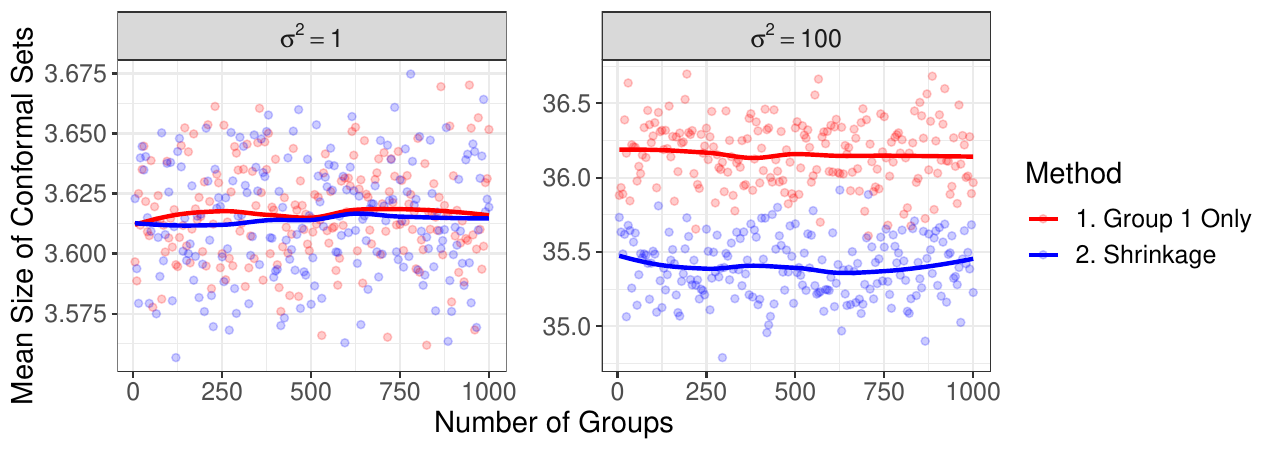}
\caption{Average size of conformal sets for a new observation from an observed group. Loess smoothing for visualization. At $\sigma^2 = 100$, within-group variance is high relative to between-group variance, and a shrinkage-based conformal method produces smaller sets.}
\label{fig:shrinkage_size}
\end{figure}

Figure \ref{fig:shrinkage_cov} shows the empirical coverage
at $1-\alpha = 0.9$, when $\sigma^2 = 1$ or $\sigma^2 = 100$.  The coverage is typically about 0.9,
with no clear difference between methods. Figure \ref{fig:shrinkage_size} plots the average size of
the conformal sets in both data
set-ups. When $\sigma^2 = 1$, the two methods produce sets with similar length. When $\sigma^2 = 100$, shrinkage consistently produces smaller sets than
using only group 1's observations. This shows that shrinkage is especially beneficial when the within-group variance is high, relative to the between-group variance. There does not appear to be a trend in set size as the number of groups increases.

\section{DATA EXAMPLE}
\label{section::example}

We now consider a data example from a sleep deprivation study \citep{balkin2000effects, belenky2003patterns}. This study evaluates 18 commercial vehicle drivers on a series of tests after $0, 1, 2, \ldots, 9$ nights of restriction to 3 hours of sleep. On each day, subjects take a series of reaction time tests, and the experimenters record each subject's average reaction time. The data are available in the \texttt{sleepstudy} dataset of \texttt{R}'s \texttt{lme4} package \citep{lme4}. 

We restructure the data to fit regressions that predict average sleep-deprived reaction time ($Y$) from number of days of sleep deprivation ($X_1$) and the subject's baseline (Day 0) average reaction time under their normal sleep amount ($X_2$). For each individual $j$, we observe nine triplets $(X_{1j}, X_{2j}, Y_j)$. For the purpose of this demonstration, we treat each $(X_{1j}, X_{2j}, Y_j)$ as a random draw from a subject-specific distribution $P_j$. (Alternatively, we could treat $X_{1j}$ as fixed, $X_{2j}$ as random, and $Y_j$ as a random draw from $P_{j, Y\mid X}$. These methods are valid as long as the nonconformity scores are exchangeable, as discussed below.) The variable $X_{1j}$ ranges from 1 to 9 days, and the baseline time $X_{2j}$ is measured once for each subject $j$. Across subjects, $X_2$ ranges from 199 to 322 milliseconds, and $Y$ ranges from 194 to 466 milliseconds. Our fitted regression models have the form 
$\hat{Y} = \hat{\beta}_1 X_{1} + \hat{\beta}_2 X_{2}.$
We have also considered a model that includes an intercept. This does not make much of a difference when assessing whether the residuals appear to be exchangeable.

Suppose we observe $(X_1, X_2) = \mathbf{x}$ on a nineteenth individual, and we want to predict the associated $Y$. We construct prediction sets $C(X_1, X_2; \alpha)$ such that $\probnew(Y\in C(X_1, X_2; \alpha)) \geq 1-\alpha$. We use the constructions from Section~\ref{section::sup_methods}, and we use nonconformity scores of $R_i(\mathbf{x},y) = |Y_i - \hat{Y}_i|$. CDF pooling uses the process justified by Theorem~\ref{thm:sup_pool}. We fit the regression model $\hat{\mu}$ on the pooled observations of 9 of the 18 individuals, and we estimate the quantiles from the remaining individuals. Single subsampling randomly selects one observation per individual. We augment the subsample with $(X_1, X_2, y)$ for the observed $(X_1, X_2)$ and some proposed $y$. We fit the regression model on this augmented sample of size 19. Repeated subsampling averages $p$-values across $B = 100$ repetitions of single subsampling using the same $(X_1, X_2, y)$.  CDF pooling is asymptotically valid ($k\to\infty$) if the nonconformity scores are exchangeable across all observations used for quantile estimation and if $Y$ is continuous. The subsampling methods are valid if the nonconformity scores are exchangeable for all subsamples of one observation per subject. From visual inspection (not shown), the CDF pooling exchangeability assumption may not be met. Several subjects have particularly high or particularly low absolute residuals on all of their observations. Also, we have $k=18$ subjects, and CDF pooling is an asymptotic method. The subsampling methods' exchangeability assumption is reasonable, based on plots of the absolute residuals when we fit and evaluate the model on one observation per subject.

\begin{figure}[H]
\centering
\includegraphics[scale=0.58]{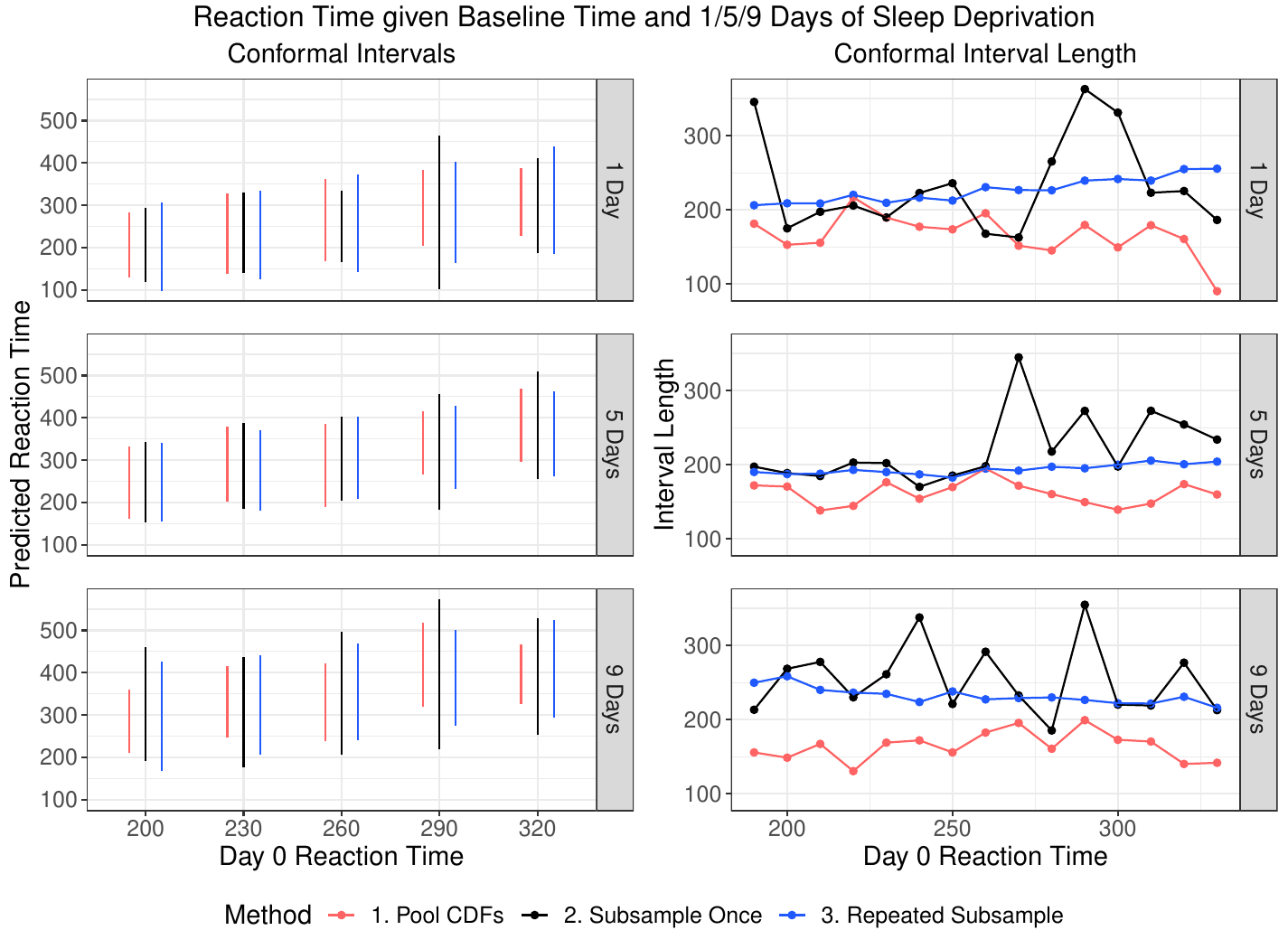}
\caption{Prediction sets for sleep-deprived reaction time given baseline reaction time and days of sleep deprivation. All sets have similar centers. CDF pooling has the smallest sets. Repeated subsampling has less variation in set size than single subsampling.}
\label{fig:sleep}
\end{figure}

Figure \ref{fig:sleep} shows the prediction sets and their size at $\alpha = 0.10$. The left panel shows the sets at $X_1 = \{1, 5, 9\}$ and at $X_2 = \{200, 230, 260, 290, 320\}$. For most $(X_1, X_2)$ combinations, all three sets have similar centers. Interestingly, the Day 1/5 prediction sets and most Day 9 prediction sets contain the Day 0 reaction time. This suggests it is plausible to maintain the baseline reaction time despite sleep deprivation. The right panel compares the length of the three sets over an expanded set of $(X_1, X_2)$ combinations. CDF pooling produces the smallest sets in most cases, but this method is only asymptotically valid ($k\to\infty$). Repeated subsampling produces smaller sets than single subsampling in about half of the cases, and it has the least variation in set lengths across $X_2$ for a given $X_1$.

We also explore the coverage of these methods. CDF pooling is only asymptotically valid ($k\to\infty$) at level $1-\alpha$, single subsampling is valid at level $1-\alpha$ but has more variation, and repeated subsampling only has guaranteed coverage at level $1-2\alpha$. We evaluate coverage by holding out 1 of the 18 individuals, selecting a triplet $(X_1, X_2, Y)$ from the held-out individual, fitting a prediction set $C$ on the remaining 17 subjects, and checking whether $Y\in C(X_1, X_2; \alpha)$. We perform this procedure $18\times 9 = 162$ times, using each observation as the test $(X_1, X_2, Y)$ once. The proportion of simulations in which $Y\in C(X_1, X_2; \alpha)$ is an estimate of the coverage. CDF pooling has algorithmic randomness in the individuals selected for model fitting (8 individuals) versus quantile estimation (9 individuals). The subsampling methods have algorithmic randomness in the observations selected for each subsample. Thus, we repeat this coverage estimation procedure 1000 times. 

Table~\ref{tab:sleep} shows the coverage proportions at $\alpha \in \{0.10, 0.15, 0.20\}$. For each method, Table~\ref{tab:sleep} displays the average coverage, the $2.5^{th}$ percentile, and the $97.5^{th}$ percentile over 1000 simulations. On average, CDF pooling undercovers by about 0.02 to 0.03, and the subsampling methods overcover by about 0.03 to 0.06. Compared to single subsampling, repeated subsampling has slightly higher coverage but lower variation in coverage. Overall, repeated subsampling is the best choice in this setting. This method achieves coverage of at least $1-\alpha$ and has lower variation in set size and coverage than the other two methods.

\begin{table}[H]
\centering
\caption{Estimated average and \mbox{($2.5^{th}$ \%ile, $97.5^{th}$ \%ile)} coverage over 1000 simulations on sleep data. CDF pooling slightly undercovers, and subsampling slightly overcovers.}
\begin{tabular}{llll}
  \hline \hline
Method & $1 - \alpha = 0.90$ & $1 - \alpha = 0.85$ &  $1 - \alpha = 0.80$ \\ \hline
1. CDF Pooling & 0.87 (0.84, 0.90) & 0.83 (0.80, 0.86) & 0.78 (0.75, 0.81) \\ 
2. Subsample Once & 0.94 (0.92, 0.97) & 0.89 (0.86, 0.92) & 0.83 (0.80, 0.87)  \\
3. Repeated Subsample & 0.95 (0.94, 0.96) & 0.91 (0.90, 0.92) & 0.84 (0.83, 0.85) \\ \hline
\end{tabular}
\label{tab:sleep}
\end{table}

\section{CONCLUSION}
\label{section::conclusion}

We have proposed and compared several methods
for constructing distribution-free prediction sets
for two-layer hierarchical models.
We believe these are the first such methods.
We consider a CDF pooling method that is asymptotically valid as $k\to\infty$, a single subsample method that uses one observation per group, and a repeated subsample method that repeatedly selects one observation per group and averages $p$-values over subsamples. The single subsample method is valid at level $1-\alpha$. The repeated subsample method has guaranteed coverage at level $1-2\alpha$ but tends to have coverage of at least $1-\alpha$ in practice.

Based on our simulations and data example, we recommend CDF pooling if asymptotic coverage is acceptable and the outcome is continuous. CDF pooling typically has the smallest prediction sets, and it yields approximately nominal coverage, especially for large $k$ and balanced groups. For small to moderate $k$ or for unbalanced groups, CDF pooling may undercover. If we desire finite sample coverage guarantees, we recommend repeated subsampling. While this method guarantees $1-2\alpha$ coverage, it achieves coverage close to $1-\alpha$ in simulations.  In the sleep example, this method has coverage of at least $1-\alpha$ and has more stable size and coverage than the other methods. Single subsampling is valid at level $1-\alpha$ but ignores most of the data. Repeated subsampling has less algorithmic variation than single subsampling, which makes this method more stable and more reproducible. It is a curiosity that single subsampling often produces slightly smaller prediction sets than repeated subsampling. 
The asymptotic efficiency of these methods relative to an oracle model remains an open question. In fact, characterizing the asymptotic efficiency of conformal methods is an open question in conformal research more broadly, with some results under additional assumptions in \cite{lei2014distribution}.

The main focus of this paper has been
the prediction of a new observation on a new subject.
In the unsupervised setting,
we also considered prediction of a future observation
on an existing subject.
Future work may consider alternatives to the James-Stein shrinkage residual or may incorporate repeated subsampling into the shrinkage approach. In addition, supervised conformal methods that borrow strength across subjects to construct prediction sets for an existing subject remain an open problem.
Space does not permit a thorough investigation of these problems,
but we hope to report more on them
in a future paper.

\section*{SUPPLEMENTARY MATERIALS}

\begin{description}

\item[Appendix~\ref{appendix:math} Mathematical Details:] This appendix (PDF file) provides proofs of Theorems \ref{thm:unsup_basic}, \ref{thm:unsup_double}, \ref{thm:unsup_pool}, \ref{thm:unsup_rep}, \ref{thm:sup_pool}, and \ref{thm:sup_model_avg}. Theorems \ref{thm::basic}, \ref{thm:unsup_once}, \ref{thm:sup_once}, \ref{thm:sup_rep}, \ref{thm:obs_isolate}, and \ref{thm:obs_shrinkage} are justified in the in-text citations. 

\item[Appendix~\ref{appendix:sims} Additional Simulations:] This appendix (PDF file) provides simulations for additional values of $n_j$ (supervised and unsupervised), non-normal data (unsupervised), and additional $(\mu, \tau^2)$ parameters (supervised).

\end{description}

\section*{ACKNOWLEDGMENTS}

This work was conducted while RD was at Carnegie Mellon University. RD's research was supported by the National Science Foundation Graduate Research Fellowship Program under Grant Nos.\ DGE 1252522 and DGE 1745016. Any opinions, findings, and conclusions or recommendations expressed in this material are those of the authors and do not necessarily reflect the views of the National Science Foundation. This work used the Extreme Science and Engineering Discovery Environment (XSEDE) \citep{xsede}, which is supported by National Science Foundation grant number ACI-1548562. Specifically, it used the Bridges system \citep{xsedebridges}, which is supported by NSF award number ACI-1445606, at the Pittsburgh Supercomputing Center (PSC). This work made extensive use of the \texttt{R} statistical software \citep{Rcore}, as well as the \texttt{data.table} \citep{datatable}, \texttt{formula.tools} \citep{formulatools},  \texttt{gridExtra} \citep{gridExtra}, \texttt{lme4} \citep{lme4}, \texttt{progress} \citep{progress}, \texttt{R.utils} \citep{Rutils}, and \texttt{tidyverse} \citep{tidyverse} packages. The authors thank the associate editor and reviewers for helpful feedback that has greatly improved the quality of the paper. The authors also thank Jing Lei and Mauricio Sadinle for helpful discussions.

\bibliographystyle{apalike}
\bibliography{paper}

\newpage

\appendix
\setcounter{page}{1}
\section{MATHEMATICAL DETAILS}
\label{appendix:math}

We recall \textbf{Theorem~\ref{thm:unsup_basic}}. Let $Y_1,\ldots, Y_n \in \mathcal{Y}$ be iid observations from a distribution $P$, where $\mathcal{Y} \subseteq \real$ or, more generally, $\mathcal{Y}$ is a linearly ordered set.
Let $Y_{n+1}$ denote a new draw from $P$.

\thmUnsupBasic*

\begin{proof}[Proof of Theorem~\ref{thm:unsup_basic}]
Suppose the data arise from a continuous distribution such that ties occur with probability 0. (This is helpful for intuition, but the inequalities that follow are valid without this assumption.) We can define the following $n+1$ sets: $$\underbrace{(\min\{\mathcal{Y}\}, Y_{(1)}]}_{A_0}, \: \underbrace{(Y_{(1)}, Y_{(2)}]}_{A_1},  \: \ldots, \: \underbrace{(Y_{(n-1)}, Y_{(n)}]}_{A_{n-1}}, \: \underbrace{(Y_{(n)}, \max\{{\mathcal{Y}}\})}_{A_n}.$$

A new observation $Y_{n+1}\sim P$ is equally likely to fall in any of those $n+1$ sets. To see this, consider the augmented sample $(Y_1, Y_2, \ldots, Y_n, Y_{n+1})$ with updated order statistics $(Y_{(1)}', Y_{(2)}', \ldots, Y_{(n)}', Y_{(n+1)}')$. The new observation $Y_{n+1}$ is equally likely to be any of those order statistics. That means
\begin{align*}P(Y_{n+1} \in A_0) &= P(Y_{n+1} \leq Y_{(1)})
= P(Y_{n+1} = Y_{(1)}')
= 1 / (n+1), \\
P(Y_{n+1} \in A_1) &= P(Y_{(1)} < Y_{n+1} \leq Y_{(2)})
= P(Y_{n+1} = Y_{(2)}')
= 1 / (n+1),
\end{align*}
and so forth. Allowing for ties, for $m \in \{1, \ldots, n\}$ we have 
\begin{align*}
    P(Y_{n+1} < Y_{(m)}) &\leq \frac{m}{n+1} \\
    P(Y_{n+1} > Y_{(m)}) &\leq \frac{n-m+1}{n+1}.
\end{align*}
These inequalities are equalities when $P$ is a continuous distribution. 
%Any union of $\lceil (1-\alpha) (n+1)\rceil$ unique sets from the collection $\{A_i\}_{i=0}^{n}$ will have $100(1-\alpha)\%$ coverage. Let $\Lambda$ contain the indices of $\lceil (1-\alpha) (n+1)\rceil$ unique sets from $\{A_i\}_{i=0}^{n}$. Allowing for tied observations, we see
%\begin{align*}
%\P\left(Y_{n+1} \in \bigcup_{\lambda \in \Lambda} A_\lambda \right) &\geq \lceil (1-\alpha) (n+1)\rceil \cdot 1 / (n+1) \geq 1-\alpha.
%\end{align*}

We construct a prediction set $[Y_{(r)}, Y_{(s)}]$ where $r = \lfloor (n+1)(\alpha/2) \rfloor$ and $s=\lceil (n+1) (1 -\alpha/2) \rceil$. We see that
\begin{align*}
P(Y_{n+1} \notin [Y_{(r)}, Y_{(s)}]) &= P(Y_{n+1} < Y_{(r)}) + P(Y_{n+1} > Y_{(s)}) \\
&\leq \frac{r}{n+1} + \frac{n-s+1}{n+1} \\
&= 1 + \frac{r-s}{n+1} \\
&= 1 + \frac{\lfloor (n+1)(\alpha/2) \rfloor - \lceil (n+1)(1-\alpha/2) \rceil}{n+1} \\
&= 1 + \frac{\lfloor (n+1)(\alpha/2) \rfloor - \left( (n+1) + \lceil -(n+1)(\alpha/2) \rceil \right)}{n+1} \\
&= \frac{\lfloor (n+1)(\alpha/2) \rfloor - \lceil -(n+1)(\alpha/2) \rceil}{n+1} \\
&= \frac{\lfloor (n+1)(\alpha/2) \rfloor + \lfloor (n+1)(\alpha/2) \rfloor}{n+1} \\
&\leq \frac{(n+1)\alpha}{n+1} \\
&= \alpha.
\end{align*}
So $P(Y_{n+1} \in [Y_{(r)}, Y_{(s)}]) \geq 1-\alpha$.

If $n \geq 2 / \alpha - 1$, then the lower bound is greater than or equal to $Y_{(1)}$ because
\begin{align*}
\lfloor (n+1)(\alpha / 2) \rfloor &\geq \lfloor \left(2/\alpha  \right) (\alpha/2) \rfloor \\
&= 1.
\end{align*}
Applying this result, we also see that the upper bound is less than or equal to $Y_{(n)}$ because
\begin{align*}
\lceil (n+1) \left(1 - \alpha / 2\right)\rceil &= \lceil n + 1 - (n+1) (\alpha / 2) \rceil \\
&= n + 1 + \lceil - (n+1) (\alpha / 2) \rceil \\
&= n + 1 - \lfloor (n+1) (\alpha / 2) \rfloor \\
&\leq n + 1 - 1 \\
&= n.
\end{align*}
\end{proof}

We recall \textbf{Theorem~\ref{thm:unsup_double}}. The data come in groups
${\cal D}_1, \ldots, {\cal D}_k$
and each group has iid data
$$
{\cal D}_j = \{ Y_{j1},\ldots, Y_{j n_j}\} \sim P_j
$$
where
$P_1,\ldots, P_k \sim \Pi$.
For this method, we assume $n_1 = n_2 = \cdots = n_k.$
(If the samples are not equally sized, then we could work with $\min_j n_j$ observations per group, sampled uniformly at random without replacement.)
Assuming a new distribution $P_{k+1}\sim \Pi$ and $Y\sim P_{k+1}$,
we want a prediction region for $Y$. We say $Y \sim \tilde{\Pi}$, where $\tilde{\Pi} = \int P d\Pi(P)$.

At the group level, let $C_j(\alpha/2) = [\ell_j, u_j]$ be the $100(1-\alpha/2)\%$ prediction set obtained by applying the method in Theorem~\ref{thm:unsup_basic} at level $\alpha/2$ to group $j$, $j = 1,\ldots, k$. We construct a vector of $k$ lower bounds $(\ell_1, \ldots, \ell_k)$ and $k$ upper bounds $(u_1, \ldots, u_k)$. Using the order statistics from those samples, we set $C^\double(\alpha) = [\ell_{(r)}, u_{(s)}]$, where $r = \lfloor (k+1)(\alpha/4) \rfloor$ and $s=\lceil (k+1)(1-\alpha/4) \rceil$. If $r < 1$ and $s > k$, let $\ell_{(r)} = \min\{\mathcal{Y}\}$ and $u_{(s)} = \max\{\mathcal{Y}\}$. 

\thmUnsupDouble*

\begin{proof}[Proof of Theorem~\ref{thm:unsup_double}]
Drawing $Y \sim \tilde{\Pi}$ is equivalent to drawing $P_{k+1}\sim\Pi$ and $Y\sim P_{k+1}$. As a helpful construct for this proof, suppose $Y_{k+1} = (Y_{k+1,1}, Y_{k+1,2}, \ldots, Y_{k+1,n_1})$ is an unobserved sample of $n_1$ additional observations from $P_{k+1}$. While the randomness in $Y$ and $C^\double(\alpha)$ are captured by $\probnew$, for the proof we define a new probability distribution $\probnewaug$. $\probnewaug$ captures the randomness over the original sample, as well as the new $Y$ and $Y_{k+1} = (Y_{k+1,1}, Y_{k+1,2}, \ldots, Y_{k+1,n_1})$. More formally, for $y\in\mathcal{Y}$ and $y_j \in \mathcal{Y}^{n_1}$, $j=1,\ldots,k+1$, we define $\probnewaug$ as
\begin{align*}
\probnewaug&(y, y_1, y_2, \ldots, y_k, y_{k+1}) \\
&= \left\{ \int P(Y \leq y) \left[\prod_{i=1}^{n_1} P(Y_{k+1,i} \leq y_{k+1,i}) \right]  d\Pi(P) \right\} \left\{ \prod_{j=1}^k \left[\int \prod_{i=1}^{n_1} P(Y_{ji} \leq y_{ji}) d\Pi(P) \right]\right\}.    
\end{align*}
Let $C_{k+1}(\alpha/2) = [\ell_{k+1}, u_{k+1}]$ be the $100(1-\alpha/2)\%$ prediction set applied to $Y_{k+1,1}, Y_{k+1,2}, \ldots, Y_{k+1,n_1}$. The $k+1$ lower bounds $\ell_1, \ldots, \ell_k, \ell_{k+1}$ are an iid sample, since we generated each value by drawing $P_j \sim \Pi$, drawing $n_1$ observations from $P_j$, and computing the $\lfloor (n_1+1)(\alpha/4)\rfloor$ order statistic of the sample.
Thus, similar to the proof of Theorem~\ref{thm:unsup_basic}, we can see that 
\begin{align*}
\probnewaug(\ell_{k+1} < \ell_{(r)}) &\leq \lfloor (k+1)(\alpha/4)\rfloor \frac{1}{k+1} \\
&\leq \alpha/4
\end{align*}
and
\begin{align*}
\probnewaug(u_{k+1} > u_{(r)}) &\leq \frac{k - \lceil (k+1)(1-\alpha/4) \rceil + 1}{k+1} \\
&\leq \alpha/4.
\end{align*}
This implies that 
\begin{align}
\probnewaug(C_{k+1}(\alpha/2) \nsubseteq C^\double(\alpha)) &= \probnewaug(\ell_{k+1} < \ell_{(r)} \cup u_{k+1} > u_{(s)}) \nonumber \\
&\leq \probnewaug(\ell_{k+1} < \ell_{(r)}) + \probnewaug(u_{k+1} > u_{(s)}) \nonumber \\
&\leq \alpha/4 + \alpha/4 \nonumber \\
&= \alpha/2. \label{eq:c_kplus1}
\end{align}

Let $A$ denote the event that $C_{k+1}(\alpha/2) \subseteq C^\double(\alpha)$. We can now show the main result:
\begin{align}
\overline{\Pi}(Y\notin C^\double(\alpha)) &= \probnewaug(Y\notin C^\double(\alpha)) \\
&= \probnewaug(Y\notin C^\double(\alpha), A) + \probnewaug(Y\notin C^\double(\alpha), A^c) \label{eq:dbl_1} \\ 
&\leq \probnewaug(Y\notin C_{k+1}(\alpha/2)) + \probnewaug(A^c) \label{eq:dbl_2} \\
&\leq \alpha/2 + \alpha/2 \label{eq:dbl_3} \\
&= \alpha. \nonumber
\end{align}
To get from (\ref{eq:dbl_1}) to (\ref{eq:dbl_2}), we note that if $Y\notin C^\double(\alpha)$ and $C_{k+1}(\alpha/2) \subseteq C^\double(\alpha)$, then $Y\notin C_{k+1}(\alpha/2)$. We also note that $\probnewaug(Y\notin C^\double(\alpha), A^c) \leq \probnewaug(A^c)$. To get from (\ref{eq:dbl_2}) to (\ref{eq:dbl_3}), the first probability uses the fact that $C_{k+1}(\alpha/2)$ was constructed as a $100(1-\alpha/2)\%$ prediction set for group $k+1$, from which $Y$ was also drawn. The second probability holds because $\probnewaug(A^c) = \probnewaug(C_{k+1}(\alpha/2) \nsubseteq C^\double(\alpha)) \leq \alpha/2$ from (\ref{eq:c_kplus1}).

Furthermore, $C^\double(\alpha)$ can produce nontrivial sets if $k\geq 4/\alpha - 1$ and $n_1 \geq 4/\alpha - 1$. We use $n_1$ observations to construct each of the sets $C_1(\alpha/2), \ldots, C_k(\alpha/2)$. Thus, by a similar argument as Theorem~\ref{thm:unsup_basic}, these $k$ sets may be nontrivial  if $n_1\geq 4/\alpha - 1$. In addition, since $C^\double(\alpha)$ uses the $\lfloor (k+1)(\alpha/4)\rfloor$ order statistic of $(\ell_1, \ldots, \ell_k)$ and the $\lceil (k+1)(1-\alpha/4)\rceil$ order statistic of $(u_1, \ldots, u_k)$, $C^\double(\alpha)$ may be nontrivial if $k\geq 4/\alpha - 1$.
\end{proof}

The unsupervised pooling method in \textbf{Theorem~\ref{thm:unsup_pool}} pools the empirical CDFs across the $k$ groups. For any group $j$ with observations $Y_{j1}, \ldots, Y_{jn_j}$, the empirical CDF is defined as $$\hat{F}_j(t) = \frac{1}{n_j} \sum_{i=1}^{n_j} I(Y_{ji} \leq t).$$ We set 
\begin{align*}
\hat{q}_k(\alpha) &= \inf \left\{ t\in\mathcal{Y} : \frac{1}{k} \sum_{j=1}^k \hat{F}_j(t) \geq \alpha \right\}.
\end{align*}
Then $C^\poolCDF(\alpha) = [\hat{q}_k(\alpha/2), \hat{q}_k(1-\alpha/2)].$

\thmUnsupPool*

\begin{proof}
Let $F(t) = \tilde{\Pi}(Y\leq t)$ and $\hat{G}_k(t) = (1/k) \sum_{j=1}^k \hat{F}_j(t)$. That means that for $\alpha\in (0,1)$, the sample quantiles $\hat{q}_k(\alpha)$ and the true quantiles $q(\alpha)$ are
\begin{align*}
\hat{q}_k(\alpha) &= \inf\left\{t\in\mathcal{Y}: \hat{G}_k(t) \geq \alpha \right\} \\
q(\alpha) &= \inf\left\{t\in\mathcal{Y}: F(t) \geq \alpha \right\}.
\end{align*}
We prove this theorem in three steps:
\begin{enumerate}
    \item For $t\in\mathcal{Y}$, $\hat{G}_k(t) \overset{p}{\to} F(t)$ as $k\to\infty$.
    \item For $\alpha\in (0,1)$, $\hat{q}_k(\alpha) \overset{p}{\to} q(\alpha)$ as $k\to\infty$.
    \item For $Y\sim\tilde{\Pi}$, $\overline{\Pi}\left(Y \in C^\poolCDF(\alpha) \right) \to 1-\alpha$ as $k\to\infty$.
\end{enumerate}

\textit{Step 1.} Fix $t\in\mathcal{Y}$. We write $$\hat{G}_k(t) = \frac{1}{k} \sum_{j=1}^k \hat{F}_j(t) = \frac{1}{k} \sum_{j=1}^k \frac{1}{n_j} \sum_{i=1}^{n_j} I(Y_{ji} \leq t).$$ We see that $\E_{\overline{\Pi}} [I(Y_{ji} \leq t)] = F(t)$, so $$\E_{\overline{\Pi}} [\hat{G}_k(t)] = F(t).$$ In addition, note that the $k$ distributions $P_1, \ldots, P_k$ are independently drawn from $\Pi$. Since $\hat{F}_j(t)$ is bounded between 0 and 1, we determine $$\text{Var}_{\overline{\Pi}}(\hat{G}_k(t)) = \text{Var}_{\overline{\Pi}}\left(\frac{1}{k} \sum_{j=1}^k \hat{F}_j(t) \right) = \frac{1}{k^2}\sum_{j=1}^k \text{Var}_{\overline{\Pi}}\left(\hat{F}_j(t)\right) \leq \frac{1}{k^2}\sum_{j=1}^k \frac{1}{4} = \frac{1}{4k} \to 0$$ as $k\to\infty$. We conclude that $\hat{G}_k(t) \overset{p}{\to} F(t)$ as $k\to\infty$.

\textit{Step 2.} Fix $\alpha\in (0, 1).$ Let $\epsilon > 0$ and $\delta > 0$. To show that $\displaystyle\lim_{k\to\infty} \overline{\Pi}\left(\left| \hat{q}_k(\alpha) - q(\alpha)\right| > \epsilon \right) = 0$, we will show that there exists $K\in\mathbb{N}$ such that for $k\geq K$, $\overline{\Pi}\left(\left| \hat{q}_k(\alpha) - q(\alpha)\right| > \epsilon \right) < \delta$.

Since we are assuming $Y$ is continuous, $F(t)$ is strictly increasing. Thus, $F(q(\alpha) + \epsilon) - F(q(\alpha)) > 0.$ Since $\hat{G}_k(t) \overset{p}{\to} F(t)$ as $k\to\infty$, we can fix $K_1\in\mathbb{N}$ such that for $k\geq K_1$, $$\overline{\Pi}\left(\left|\hat{G}_k(q(\alpha) + \epsilon) - F(q(\alpha) + \epsilon) \right| > F(q(\alpha) + \epsilon) - F(q(\alpha)) \right) < \frac{\delta}{2}.$$ In addition, $F(q(\alpha)) - F(q(\alpha) - \epsilon) > 0$. We can fix $K_2\in\mathbb{N}$ such that for $k\geq K_2$, $$\overline{\Pi}\left(\left|\hat{G}_k(q(\alpha) - \epsilon) - F(q(\alpha) - \epsilon) \right| \geq F(q(\alpha)) - F(q(\alpha) - \epsilon) \right) < \frac{\delta}{2}.$$ Now let $K = \max\{K_1, K_2\}$. Assume $k\geq K$. From the definition of $\hat{q}_k(\alpha)$, it holds that $\hat{q}_k(\alpha) > q(\alpha) + \epsilon$ implies $\hat{G}_k(q(\alpha) + \epsilon) < \alpha$. It also holds that $\hat{q}_k(\alpha) < q(\alpha) - \epsilon$ implies $\hat{G}_k(q(\alpha) - \epsilon) \geq \alpha$. We see
\begin{align*}
\overline{\Pi}&\left(\left|\hat{q}_k(\alpha) - q(\alpha) \right| > \epsilon \right) \\
&= \overline{\Pi}\left(\hat{q}_k(\alpha) > q(\alpha) + \epsilon \right) + \overline{\Pi}(\hat{q}_k(\alpha) < q(\alpha) - \epsilon) \\
&\leq \overline{\Pi}\left(\hat{G}_k(q(\alpha) + \epsilon) < \alpha \right) + \overline{\Pi}\left(\hat{G}_k(q(\alpha) - \epsilon) \geq \alpha \right) \\
&= \overline{\Pi}\left(\hat{G}_k(q(\alpha) + \epsilon) < F(q(\alpha)) \right) + \overline{\Pi}\left(\hat{G}_k(q(\alpha) - \epsilon) \geq F(q(\alpha)) \right) \addtocounter{equation}{1}\tag{\theequation} \label{eq:cts_inc} \\ 
&= \overline{\Pi}\left(\hat{G}_k(q(\alpha) + \epsilon) < F(q(\alpha) + \epsilon) - (F(q(\alpha) + \epsilon) - F(q(\alpha))) \right) \\
&\qquad + \overline{\Pi}\left(\hat{G}_k(q(\alpha) - \epsilon) \geq F(q(\alpha) - \epsilon) + (F(q(\alpha)) - F(q(\alpha) - \epsilon)) \right) \\
&\leq \overline{\Pi}\left(\left|\hat{G}_k(q(\alpha) + \epsilon) - F(q(\alpha) + \epsilon) \right| > F(q(\alpha) + \epsilon) - F(q(\alpha)) \right) \\
&\qquad + \overline{\Pi}\left(\left|\hat{G}_k(q(\alpha) - \epsilon) - F(q(\alpha) - \epsilon) \right| \geq F(q(\alpha)) - F(q(\alpha) - \epsilon) \right) \\
&< \frac{\delta}{2} + \frac{\delta}{2} \\
&= \delta.
\end{align*}
Line (\ref{eq:cts_inc}) uses the property that $F(t)$ is continuous and strictly increasing since $Y$ follows a continuous distribution. We conclude that $\hat{q}_k(\alpha) \overset{p}{\to} q(\alpha)$ as $k\to\infty$.

\textit{Step 3.} Now we show that $\lim_{k\to\infty} \overline{\Pi}\left(Y \in C^\poolCDF(\alpha)\right) = 1-\alpha$. We see 
\begin{align}
\lim_{k\to\infty} \overline{\Pi}\left(Y \in C^\poolCDF(\alpha)\right) &= \lim_{k\to\infty} \Big[ \overline{\Pi}(Y \leq \hat{q}_k(1-\alpha/2)) - \overline{\Pi}(Y \leq \hat{q}_k(\alpha/2)) \Big]. \label{eq:thm4_step3_init}
\end{align}
We consider $\lim_{k\to\infty} \overline{\Pi}(Y \leq \hat{q}_k(1-\alpha/2))$. Fix a new $\epsilon > 0$. Then
\begin{align*}
\lim_{k\to\infty} &\overline{\Pi}(Y \leq \hat{q}_k(1-\alpha/2)) \\
&= \lim_{k\to\infty} \overline{\Pi}(Y \leq \hat{q}_k(1-\alpha/2) - q(1-\alpha/2) + q(1-\alpha/2)) \\
&= \lim_{k\to\infty} \Big[ \overline{\Pi}(Y \leq \hat{q}_k(1-\alpha/2) - q(1-\alpha/2) + q(1-\alpha/2), \hat{q}_k(1-\alpha/2) - q(1-\alpha/2) < \epsilon) + \\
&\hspace*{4em} \overline{\Pi}(Y \leq \hat{q}_k(1-\alpha/2) - q(1-\alpha/2) + q(1-\alpha/2), \hat{q}_k(1-\alpha/2) - q(1-\alpha/2) \geq \epsilon) \Big] \\
&\leq \lim_{k\to\infty} \overline{\Pi}(Y \leq q(1-\alpha/2) + \epsilon) + \lim_{k\to\infty} \overline{\Pi}(\hat{q}_k(1-\alpha/2) - q(1-\alpha/2) \geq \epsilon) \\
&\leq \tilde{\Pi}(Y \leq q(1-\alpha/2) + \epsilon) + \lim_{k\to\infty} \overline{\Pi}(|\hat{q}_k(1-\alpha/2) - q(1-\alpha/2)| \geq \epsilon) \\
&= \tilde{\Pi}(Y \leq q(1-\alpha/2) + \epsilon).
\end{align*}

In addition,
\begin{align*}
\lim_{k\to\infty} &\overline{\Pi}(Y \leq \hat{q}_k(1-\alpha/2)) \\
&= \lim_{k\to\infty} \overline{\Pi}(Y \leq \hat{q}_k(1-\alpha/2) - q(1-\alpha/2) + q(1-\alpha/2)) \\
&= \lim_{k\to\infty} \Big[ \overline{\Pi}(Y \leq \hat{q}_k(1-\alpha/2) - q(1-\alpha/2) + q(1-\alpha/2),  q(1-\alpha/2) - \hat{q}_k(1-\alpha/2) < \epsilon) + \\
&\hspace*{4em} \overline{\Pi}(Y \leq \hat{q}_k(1-\alpha/2) - q(1-\alpha/2) + q(1-\alpha/2), q(1-\alpha/2) - \hat{q}_k(1-\alpha/2)\geq \epsilon) \Big].
\end{align*}
The limit of the second expression is 0 because
\begin{align*}
\lim_{k\to\infty} &\overline{\Pi}(Y \leq \hat{q}_k(1-\alpha/2) - q(1-\alpha/2) + q(1-\alpha/2), q(1-\alpha/2) - \hat{q}_k(1-\alpha/2)\geq \epsilon)  \\
&\leq \lim_{k\to\infty} \overline{\Pi}(q(1-\alpha/2) - \hat{q}_k(1-\alpha/2)\geq \epsilon) \\
&\leq \lim_{k\to\infty} \overline{\Pi}(|q(1-\alpha/2) - \hat{q}_k(1-\alpha/2)|\geq \epsilon) \\
&= 0.
\end{align*}
So 
\begin{align*}
\lim_{k\to\infty} &\overline{\Pi}(Y \leq \hat{q}_k(1-\alpha/2)) \\
&= \lim_{k\to\infty} \overline{\Pi}(Y \leq \hat{q}_k(1-\alpha/2) - q(1-\alpha/2) + q(1-\alpha/2),  \: q(1-\alpha/2) - \hat{q}_k(1-\alpha/2) < \epsilon) \\
&\geq \lim_{k\to\infty} \overline{\Pi}(Y \leq - \epsilon + q(1-\alpha/2),\:  \hat{q}_k(1-\alpha/2) - q(1-\alpha/2) > - \epsilon) \\
&= \tilde{\Pi}(Y \leq - \epsilon + q(1-\alpha/2)) \left\{\lim_{k\to\infty} \overline{\Pi}(\hat{q}_k(1-\alpha/2) - q(1-\alpha/2) > - \epsilon) \right\}
\end{align*}
because $Y$ is independent of $\hat{q}_k(1-\alpha/2)$. We know that
\begin{align*}
\lim_{k\to\infty} \overline{\Pi}(\hat{q}_k(1-\alpha/2) - q(1-\alpha/2) > - \epsilon) &= \lim_{k\to\infty} \overline{\Pi}(q(1-\alpha/2) - \hat{q}_k(1-\alpha/2) < \epsilon)\\
&\geq  \lim_{k\to\infty} \overline{\Pi}(|q(1-\alpha/2) - \hat{q}_k(1-\alpha/2)| < \epsilon) \\
&= 1.
\end{align*}
Thus, we have shown that for arbitrary $\epsilon > 0$, 
\begin{align}
\tilde{\Pi}(Y \leq q(1-\alpha/2) - \epsilon) \leq \lim_{k\to\infty} \overline{\Pi}(Y \leq \hat{q}_k(1-\alpha/2)) \leq \tilde{\Pi}(Y \leq q(1-\alpha/2) + \epsilon).  \label{eq:thm4_ineq}
\end{align}
Then $\lim_{k\to\infty} \overline{\Pi}(Y \leq \hat{q}_k(1-\alpha/2)) = 1-\alpha/2.$ 

To see why, suppose $\lim_{k\to\infty} \overline{\Pi}(Y \leq \hat{q}_k(1-\alpha/2))  > 1-\alpha/2$. Since $\tilde{\Pi}(Y \leq t)$ is continuous and strictly increasing, there exists $\epsilon_1 > 0$ such that $\tilde{\Pi}(Y < \epsilon_1 + q(1-\alpha/2)) = \lim_{k\to\infty} \overline{\Pi}(Y \leq \hat{q}_k(1-\alpha/2))$. Then at $\epsilon_2 = \epsilon_1 / 2$, we have $$\tilde{\Pi}(Y < \epsilon_2 + q(1-\alpha/2)) < \tilde{\Pi}(Y < \epsilon_1 + q(1-\alpha/2)) = \lim_{k\to\infty} \overline{\Pi}(Y \leq \hat{q}_k(1-\alpha/2)),$$ which contradicts (\ref{eq:thm4_ineq}). Alternatively, suppose $\lim_{k\to\infty} \overline{\Pi}(Y \leq \hat{q}_k(1-\alpha/2))  < 1-\alpha/2$. Since $\tilde{\Pi}(Y \leq t)$ is continuous and strictly increasing, there exists $\epsilon_1 > 0$ such that $\tilde{\Pi}(Y < q(1-\alpha/2) - \epsilon_1) = \lim_{k\to\infty} \overline{\Pi}(Y \leq \hat{q}_k(1-\alpha/2))$. Then at $\epsilon_2 = \epsilon_1 / 2$, we have $$\tilde{\Pi}(Y < q(1-\alpha/2) - \epsilon_2) > \tilde{\Pi}(Y < q(1-\alpha/2) - \epsilon_1) = \lim_{k\to\infty} \overline{\Pi}(Y \leq \hat{q}_k(1-\alpha/2)),$$ which also contradicts (\ref{eq:thm4_ineq}).

Thus, we have shown that $\lim_{k\to\infty} \overline{\Pi}(Y \leq \hat{q}_k(1-\alpha/2)) = 1-\alpha/2.$  Returning to (\ref{eq:thm4_step3_init}), we conclude that 
\begin{align*}
\lim_{k\to\infty} \overline{\Pi}\left(Y \in C^\poolCDF(\alpha)\right) &= \lim_{k\to\infty} \Big[ \overline{\Pi}(Y \leq \hat{q}_k(1-\alpha/2)) - \overline{\Pi}(Y \leq \hat{q}_k(\alpha/2)) \Big] \\
&= (1 - \alpha/2) - \alpha/2 \\
&= 1-\alpha.
\end{align*}

The set given by $C^\poolCDF(\alpha) = [\hat{q}_k(\alpha/2), \hat{q}_k(1-\alpha/2)]$ may be nontrivial for any $k\geq 1$ and $n_j\geq 1$, $j=1,\ldots, k$. To see this, suppose $\displaystyle t_1 < \min_{1\leq j\leq k} \min_{1\leq i\leq n_j} Y_{ji}$. Then $(1/k) \sum_{j=1}^k \hat{F}_j(t_1) = 0$. This implies that $$\hat{q}_k(\alpha/2) = \inf\left\{t\in\mathcal{Y}: (1/k) \sum_{j=1}^k \hat{F}_j(t) \geq \alpha/2 \right\} > t_1.$$ Next, suppose $\displaystyle t_2 = \max_{1\leq j\leq k} \max_{1\leq i\leq n_j} Y_{ji}$. Then $(1/k) \sum_{j=1}^k \hat{F}_j(t_2) = 1$. This implies that $$\hat{q}_k(1 - \alpha/2) = \inf\left\{t\in\mathcal{Y}: (1/k) \sum_{j=1}^k \hat{F}_j(t) \geq 1 - \alpha/2 \right\} \leq t_2.$$ Thus, $C^\poolCDF(\alpha)$ is a subset of $[\min_j \{Y_{j1}, \ldots, Y_{jn_j}\}, \max_j \{Y_{j1}, \ldots, Y_{jn_j}\}]$.
\end{proof}

We recall the setup for the unsupervised repeated subsampling method of \textbf{Theorem~\ref{thm:unsup_rep}}. This method takes $B$ subsamples of a single observation from each of the $k$ groups. Suppose $Y^b_{(1)}, Y^b_{(2)}, \ldots, Y^b_{(k)}$ are the ordered observations from the $b^{th}$ subsample. Within the $b^{th}$ subsample, a valid $p$-value for the test of $H_0: Y_{k+1} = u$ versus $H_1: Y_{k+1}\neq u$ is
\[\pi_b(u) = \begin{cases} 
      1 & \text{if } u \in [Y^b_{(\lfloor (k+1)/2 \rfloor)}, Y^b_{(\lceil (k+1)/2 \rceil)}] \\
      \inf\{\alpha : u\notin [Y_{(r)}^b, Y_{(s)}^b] \} & \text{otherwise}
   \end{cases},
\]
 where $r = \lfloor (k+1)(\alpha/2) \rfloor$ and $s = \lceil (k+1)(1-\alpha/2) \rceil$. We define a prediction set $$C^\rep(\alpha) = \left\{u: \frac{1}{B}\sum_{b=1}^B \pi_b(u) \geq \alpha\right\}.$$ 

\thmUnsupRep*

\begin{proof}
As stated in the main text, $(2/B)\sum_{b=1}^B \pi_b(u)$ (double the test statistic) is a valid $p$-value for the test of $H_0: Y_{k+1} = u$ versus $H_1: Y_{k+1}\neq u$ \citep{ruschendorf1982random, meng1994posterior, barber2021predictive, vovk2020}. The set of all $u$ at which we would not reject $H_0$ at level $2\alpha$ is a valid $100(1-2\alpha)\%$ prediction set. Hence, $$C^\rep(\alpha) = \left\{u: \frac{2}{B}\sum_{b=1}^B \pi_b(u) \geq 2\alpha\right\} = \left\{u: \frac{1}{B}\sum_{b=1}^B \pi_b(u) \geq \alpha\right\}$$ is a valid  $100(1-2\alpha)\%$ prediction set. 

If $k > 2/\alpha - 1$ and each $n_j \geq 1$, then $C^\rep(\alpha)$ may be nontrivial. Suppose $\displaystyle t_1 < \min_{1\leq j\leq k} \min_{1\leq i\leq n_j} Y_{ji}$ and $\displaystyle t_2 > \max_{1\leq j\leq k} \max_{1\leq i\leq n_j} Y_{ji}$. Then for each $b$ we have $t_1 < Y^b_{(1)}$, which means $\pi_b(t_1) = 2/(k+1).$ Similarly, for each $b$ we have $t_2 > Y^b_{(k)}$, which means $\pi_b(t_2) = 2/(k+1).$ If $k > 2/\alpha - 1$, then $$\frac{1}{B} \sum_{b=1}^B \pi_b(t_1) = \frac{2}{k+1} <  \alpha$$ and $$\frac{1}{B} \sum_{b=1}^B \pi_b(t_2) = \frac{2}{k+1} < \alpha.$$ This means $t_1$ and $t_2$ are outside $C^\rep(\alpha)$, where $t_1$ is an arbitrary value less than the minimum observation, and $t_2$ is an arbitrary value greater than the maximum observation. Thus, if $k > 2/\alpha - 1$, then $C^\rep(\alpha)$ is a subset of $[\min_j \{Y_{j1}, \ldots, Y_{jn_j}\}, \max_j \{Y_{j1}, \ldots, Y_{jn_j}\}]$.
\end{proof}

We recall the setup for the supervised CDF pooling method referenced in \textbf{Theorem~\ref{thm:sup_pool}}. Let $[k] = \{1, \ldots, k\}$. We start by pooling the observations from some strict subset $k_0 \subset [k]$ of the $k$ groups to fit a model $\hat{\mu}(X)$ as an estimator of $\E[Y \mid X]$. We use the remaining groups to fit the residuals $R_{ji} = |Y_{ji} - \hat{\mu}(X_{ji})|$, $j\in [k]\backslash k_0$, $i = 1,\ldots, n_j$. Now for each $j\in [k]\backslash k_0$, we define group $j$'s empirical CDF of the residuals $$\hat{F}_j(t) = \frac{1}{n_j} \sum_{i=1}^{n_j} I(R_{ji} \leq t).$$ We define $$\hat{q}_k(\alpha) = \inf\left\{t\in\real: \frac{1}{|[k]\backslash k_0|} \sum_{j\in [k]\backslash k_0} \hat{F}_j(t) \geq \alpha \right\}.$$  The $1-\alpha$ prediction set is $C^\poolCDF(x; \alpha) = [\hat{\mu}(x) - \hat{q}_k(1-\alpha), \hat{\mu}(x) +  \hat{q}_k(1-\alpha)]$. 

\thmSupPool*

\begin{proof}
The proof of Theorem~\ref{thm:sup_pool} is similar to the proof of Theorem~\ref{thm:unsup_pool}. We explain how to modify the argument to prove the supervised result. Let $\hat{G}_k(t) = \left(|[k] \backslash k_0| \right)^{-1} \sum_{j\in [k] \backslash k_0} \hat{F}_j(t)$. Let $F(t) = \tilde{\Pi}(|Y - \hat{\mu}(X)| \leq t)$. For $\alpha\in (0,1)$, the sample quantiles $\hat{q}_k(\alpha)$ and the true quantiles $q(\alpha)$ are 
\begin{align*}
\hat{q}_k(\alpha) &= \inf\left\{t\in\real: \hat{G}_k(t) \geq \alpha \right\} \\
q(\alpha) &= \inf\left\{t\in\real: F(t) \geq \alpha \right\}.
\end{align*}
Similar to Theorem~\ref{thm:unsup_pool}, we prove this theorem in three steps:
\begin{enumerate}
  \item For $t\in\real$, $\hat{G}_k(t) \overset{p}{\to} F(t)$ as $k\to\infty$.
  \item For $\alpha\in (0,1)$, $\hat{q}_k(\alpha) \overset{p}{\to} q(\alpha)$ as $k\to\infty$.
  \item For $(X,Y)\sim\overline{\Pi}$, $\overline{\Pi}\left(Y \in C^\poolCDF(X;\alpha) \right) \to 1-\alpha$ as $k\to\infty$.
\end{enumerate}
\textit{Step 1.} Fix $t\in\real$. We write $$\hat{G}_k(t) = \frac{1}{|[k] \backslash k_0|} \sum_{j\in [k] \backslash k_0} \hat{F}_j(t) = \frac{1}{|[k] \backslash k_0|} \sum_{j\in [k] \backslash k_0} \frac{1}{n_j} \sum_{i=1}^{n_j} I(|Y_{ji} - \hat{\mu}(X_{ji})| \leq t).$$ Again, we assume that $\hat{\mu}(\cdot)$ is fixed, given the observations in the groups indexed by $k_0$. For $j\in [k] \backslash k_0$, we know $\E_{\overline{\Pi}}[I(|Y_{ji} - \hat{\mu}(X_{ji})| \leq t)] = \overline{\Pi}(|Y - \hat{\mu}(X)| \leq t)$. Hence,
\begin{align*}
\E_{\overline{\Pi}}[\hat{G}_k(t)] &= \overline{\Pi}(|Y - \hat{\mu}(X)| \leq t) = F(t).
\end{align*}
% \begin{align*}
% \E_{\overline{\Pi}}[\hat{G}_k(t) - F(t)] &= \E_{\overline{\Pi}}[ \E_{\overline{\Pi}}[ \hat{G}_k(t) - F(t) \mid (\mathcal{D}_j)_{j\in k_0}]] \\
% &= \E_{\overline{\Pi}}[ \overline{\Pi}(|Y - \hat{\mu}(X)| \geq t) - \overline{\Pi}(|Y - \hat{\mu}(X)| \geq t)] \\
% &= 0. 
% \end{align*}

Since $\hat{F}_j(t)$ is bounded between 0 and 1 and since the distributions $P_j$, $j\in [k]\backslash k_0$, are independently drawn from $\Pi$, we see
\begin{align*}
\text{Var}_{\overline{\Pi}}(\hat{G}_k(t)) &= \text{Var}_{\overline{\Pi}}\left(\frac{1}{|[k] \backslash k_0|} \sum_{j\in [k] \backslash k_0} \hat{F}_j(t)\right) \\
&= \left(\frac{1}{|[k] \backslash k_0|} \right)^2 \sum_{j\in [k] \backslash k_0} \text{Var}_{\overline{\Pi}} (\hat{F}_j(t)) \\
&\leq \frac{1}{4 |[k] \backslash k_0| } \\
&\to 0.
\end{align*}

% \begin{align*}
% \text{Var}_{\overline{\Pi}}(\hat{G}_k(t) - F(t)) &= \E_{\overline{\Pi}}\left[\text{Var}_{\overline{\Pi}}\left(\hat{G}_k(t) - F(t) \mid (\mathcal{D}_j)_{j\in k_0}\right)\right] + \text{Var}_{\overline{\Pi}}\left(\E_{\overline{\Pi}}\left[\hat{G}_k(t) - F(t) \mid (\mathcal{D}_j)_{j\in k_0} \right] \right) \\
% &= \E_{\overline{\Pi}}\left[\text{Var}_{\overline{\Pi}}\left(\hat{G}_k(t) - F(t) \mid (\mathcal{D}_j)_{j\in k_0}\right)\right] \\
% &= \E_{\overline{\Pi}}\left[\text{Var}_{\overline{\Pi}}\left(\hat{G}_k(t) \mid (\mathcal{D}_j)_{j\in k_0}\right)\right] \\
% &= \E_{\overline{\Pi}}\left[ \text{Var}_{\overline{\Pi}}\left(\frac{1}{|[k] \backslash k_0|} \sum_{j\in [k] \backslash k_0} \hat{F}_j(t) \: \Big| \: (\mathcal{D}_j)_{j\in k_0} \right) \right] \\
% &= \E_{\overline{\Pi}}\left[ \left(\frac{1}{|[k] \backslash k_0|} \right)^2 \sum_{j\in [k] \backslash k_0} \text{Var}_{\overline{\Pi}} (\hat{F}_j(t) \mid (\mathcal{D}_j)_{j\in k_0}) \right] \\
%   &\leq \frac{1}{4 |[k] \backslash k_0| } \to 0
% \end{align*}
% as $k\to\infty$. 

We conclude that for fixed $\hat{\mu}$, $\hat{G}_k(t) \overset{p}{\to} F(t)$ as $k\to\infty$.

\textit{Step 2.} We can show that for $\alpha\in (0,1)$, $\hat{q}_k(\alpha) \overset{p}{\to} q(\alpha)$ as $k\to\infty$ using the same steps as in the proof of Theorem~\ref{thm:unsup_pool}. The only modification is that $\hat{G}_k(t)$ and $F(t)$ have different definitions in the supervised case.

\textit{Step 3.} Recall that $C^\poolCDF(x;\alpha) = [\hat{\mu}(x) - \hat{q}_k(1-\alpha), \hat{\mu}(x) + \hat{q}_k(1-\alpha)]$. For $(X, Y)$ randomly drawn from $\tilde{\Pi}$, we know that
\begin{align*}
 \lim_{k\to\infty} \overline{\Pi}\left(Y \in C^\poolCDF(X;\alpha) \right) &= \lim_{k\to\infty} \overline{\Pi}\left(|Y - \hat{\mu}(X)| \leq \hat{q}_k(1-\alpha)\right).
\end{align*}
The proof of Theorem~\ref{thm:unsup_pool} considered a similar setting in the unsupervised case. We can modify step 3 of the proof of Theorem~\ref{thm:unsup_pool}, replacing $Y$ with $|Y - \hat{\mu}(X)|$, to conclude that $$\lim_{k\to\infty} \overline{\Pi}\left(|Y - \hat{\mu}(X)| \leq \hat{q}_k(1-\alpha)\right) = 1-\alpha.$$

The set $C^\poolCDF(x; \alpha) = [\hat{\mu}(x) - \hat{q}_k(1-\alpha), \hat{\mu}(x) + \hat{q}_k(1-\alpha)]$ may be nontrivial if $k\geq 1$ and each $n_j \geq 1$, $j=1,\ldots, k$. (If $k = 1$, then we would fit $\hat{\mu}(X)$ without using the data, since we fit $\hat{\mu}(X)$ using the observations in some strict subset $k_0$ of $[k]$. We would not expect the approach to have good coverage at $k=1$.) To see that the set is nontrivial, note that if $t = \displaystyle \max_{1\leq j\leq k} \max_{1\leq i\leq n_j} R_{ji} =  \max_{1\leq j\leq k} \max_{1\leq i\leq n_j} |Y_{ji} - \hat{\mu}(X_{ji})|$, then each $\hat{F}_j(t) = 1$. Thus, $\displaystyle \hat{q}_k(1-\alpha) \leq \max_{1\leq j\leq k} \max_{1\leq i\leq n_j} R_{ji}$, and the length of $C^\poolCDF(x; \alpha)$ is at most $\displaystyle 2\left(\max_{1\leq j\leq k} \max_{1\leq i\leq n_j} R_{ji}\right)$. 
\end{proof}

We recall the setup for the supervised parametric CDF pooling method referenced in \textbf{Theorem~\ref{thm:sup_model_avg}}. We also introduce some additional parameters for the proof. If $(X_j,Y_j) \sim P_j$, then suppose $Y_j = \mu_{P_j}(X_j) + \epsilon$, where  $\epsilon$ has a zero-mean distribution. For each group $j\in\{1,\ldots,k\}$, we use the $n_j$ observations $X_{j1}, X_{j2}, \ldots X_{jn_j}$ in group $j$ to fit a model $\hat{\mu}_{P_j}$. At any given $x$, we define a pooled model $\mu(x)$ and an estimated pooled model $\hat{\mu}(x)$ as 
\begin{align*}
\mu(x) &= \int \mu_P(x) d\Pi(P) \\
\hat{\mu}(x) &= k^{-1} \sum_{j=1}^k \hat{\mu}_{P_j}(x).
\end{align*}
Thus, unlike in Theorem~\ref{thm:sup_pool}, $\hat{\mu}$ changes as $k$ increases. We have the following residuals under $\mu$ and $\hat{\mu}$:
\begin{align*}
R_{ji}(\mu) &= \left|\mu(X_{ji}) - Y_{ji} \right| \\
R_{ji}(\hat{\mu}) &= \left|\hat{\mu}(X_{ji}) - Y_{ji} \right|.
\end{align*}
The empirical CDFs of these residuals are
\begin{align*}
\hat{F}_{j,\mu}(t) &= \frac{1}{n_j} \sum_{i=1}^{n_j} I(R_{ji}(\mu) \leq t) \\
\hat{F}_{j,\hat{\mu}}(t) &= \frac{1}{n_j} \sum_{i=1}^{n_j} I(R_{ji}(\hat{\mu}) \leq t).
\end{align*}  
Where $\tilde{\Pi} = \int P d\Pi(P)$, the true CDF of the residuals is
\begin{align*}
F(t) &= \tilde{\Pi}(|Y - \mu(X)| \leq t).
\end{align*}
We obtain sample quantiles $\hat{q}_k(\hat{\mu}; \alpha)$ and true quantiles $q(\alpha)$:
\begin{align*}
  \hat{q}_k(\hat{\mu}; \alpha) &= \inf\left\{t\in\mathbb{R}: \frac{1}{k}\sum_{j=1}^k \hat{F}_{j,\hat{\mu}}(t) \geq \alpha \right\} \\
  q(\alpha) &= \inf\left\{t\in\mathbb{R}: F(t) \geq \alpha \right\}.
\end{align*}
Under the assumptions stated in Theorem~\ref{thm:sup_model_avg}, an asymptotic $1-\alpha$ prediction set is $C^\modelAvg(x; \alpha) = [\hat{\mu}(x) - \hat{q}_k(\hat{\mu}; 1-\alpha), \hat{\mu}(x) + \hat{q}_k(\hat{\mu}; 1-\alpha)]$.

\thmSupModelAvg*

\begin{proof}
The proof of Theorem~\ref{thm:sup_model_avg} is similar to the proof of Theorem~\ref{thm:sup_pool}. Define 
\begin{align*}
\hat{G}_{k,\hat{\mu}}(t) &= (1/k)\sum_{j=1}^k \hat{F}_{j,\hat{\mu}}(t) \\
\hat{G}_{k,\mu}(t) &= (1/k)\sum_{j=1}^k \hat{F}_{j,\mu}(t).
\end{align*}
Similar to Theorem~\ref{thm:sup_pool}, we prove this theorem in three steps:
\begin{enumerate}
  \item For $t\in\real$, $\hat{G}_{k,\hat{\mu}}(t) \overset{p}{\to} F(t)$ as $k\to\infty$.
  \item For $\alpha\in (0,1)$, $\hat{q}_k(\hat{\mu}; \alpha) \overset{p}{\to} q(\alpha)$ as $k\to\infty$.
  \item For $(X,Y)\sim\tilde{\Pi}$, $\overline{\Pi}\left(Y \in C^\modelAvg(X;\alpha) \right) \to 1-\alpha$ as $k\to\infty$.
\end{enumerate}
\textit{Step 1.} Fix $t\in\real$. By the assumption that $(1/k)\sum_{j=1}^k \sup_t |\hat{F}_{j,\hat{\mu}}(t) - \hat{F}_{j,\mu}(t)| \overset{p}{\to} 0$ as $k\to\infty$, we know that $$|\hat{G}_{k,\hat{\mu}}(t) - \hat{G}_{k,\mu}(t)| \leq \frac{1}{k}\sum_{j=1}^k |\hat{F}_{j,\hat{\mu}}(t) - \hat{F}_{j,\mu}(t)| \overset{p}{\to} 0.$$
Next, we write $$\hat{G}_{k,\mu}(t) = \frac{1}{k} \sum_{j=1}^k \hat{F}_{j,\mu}(t) = \frac{1}{k} \sum_{j=1}^k \frac{1}{n_j} \sum_{i=1}^{n_j} I(|Y_{ji} - \mu(X_{ji})| \leq t).$$ 
We see that $\E_{\overline{\Pi}}[I(|Y_{ji} - \mu(X_{ji})| \leq t)] = \overline{\Pi}(|Y - \mu(X)| \leq t) = \tilde{\Pi}(|Y - \mu(X)| \leq t)$, so $$E_{\overline{\Pi}}[\hat{G}_{k,\mu}(t)] = \tilde{\Pi}(|Y - \mu(X)| \leq t) = F(t).$$ Since $\hat{F}_{j,\mu}(t)$ is bounded between 0 and 1, we see $$\text{Var}_{\overline{\Pi}}(\hat{G}_{k,\mu}(t)) = \text{Var}_{\overline{\Pi}}\left(\frac{1}{k} \sum_{j=1}^k \hat{F}_{j,\mu}(t)\right) = \frac{1}{k^2} \sum_{j=1}^k \text{Var}_{\overline{\Pi}} \left(\hat{F}_{j,\mu}(t)\right) \leq \frac{1}{4k} \to 0.$$ That means that $\hat{G}_{k,\mu}(t) \overset{p}{\to} F(t)$ as $k\to\infty$. Combining these two convergence statements, $$|\hat{G}_{k,\hat{\mu}}(t) - F(t)| = |\hat{G}_{k,\hat{\mu}}(t) - \hat{G}_{k,\mu}(t) + \hat{G}_{k,\mu}(t) - F(t)| \leq |\hat{G}_{k,\hat{\mu}}(t) - \hat{G}_{k,\mu}(t)| + |\hat{G}_{k,\mu}(t) - F(t)| \overset{p}{\to} 0.$$ We conclude that $\hat{G}_{k,\hat{\mu}}(t) \overset{p}{\to} F(t)$ as $k\to\infty$.

\textit{Step 2.} We can show that for $\alpha\in (0,1)$, $\hat{q}_k(\hat{\mu}; \alpha) \overset{p}{\to} q(\alpha)$ as $k\to\infty$ using the same steps as in the proof of Theorem~\ref{thm:unsup_pool}. As modifications, we replace $\hat{G}_k(t)$ with $\hat{G}_{k,\hat{\mu}}(t)$, and we use $F(t) = \tilde{\Pi}(|Y - \mu(X)| \leq t)$.

\textit{Step 3.} Recall that $C^\modelAvg(x; \alpha) = [\hat{\mu}(x) - \hat{q}_k(\hat{\mu}; 1-\alpha), \hat{\mu}(x) + \hat{q}_k(\hat{\mu}; 1-\alpha)]$. For $(X, Y)$ randomly drawn from $\tilde{\Pi}$, we know that
\begin{align*}
 \lim_{k\to\infty} \overline{\Pi}\left(Y \in C^\modelAvg(X;\alpha) \right) &= \lim_{k\to\infty} \overline{\Pi}\left(|Y - \hat{\mu}(X)| \leq \hat{q}_k(\hat{\mu}; 1-\alpha)\right).
\end{align*}
The proof of Theorem~\ref{thm:unsup_pool} considered a similar setting in the unsupervised case. We can modify step 3 of the proof of Theorem~\ref{thm:unsup_pool}, replacing $Y$ with $|Y - \hat{\mu}(X)|$ and replacing $\hat{q}_k(1-\alpha/2)$ with $\hat{q}_k(\hat{\mu}; 1-\alpha)$, to conclude that $$\lim_{k\to\infty} \overline{\Pi}\left(|Y - \hat{\mu}(X)| \leq \hat{q}_k(\hat{\mu}; 1-\alpha)\right) = 1-\alpha.$$

The set $C^\modelAvg(x; \alpha) = [\hat{\mu}(x) - \hat{q}_k(\hat{\mu}; 1-\alpha), \hat{\mu}(x) + \hat{q}_k(\hat{\mu}; 1-\alpha)]$ may be nontrivial if $k\geq 1$ and each $n_j \geq 1$, $j=1,\ldots, k$. To see this, note that if $t = \displaystyle \max_{1\leq j\leq k} \max_{1\leq i\leq n_j} R_{ji}(\hat{\mu}) =  \max_{1\leq j\leq k} \max_{1\leq i\leq n_j} |Y_{ji} - \hat{\mu}(X_{ji})|$, then $\hat{F}_{j, \hat{\mu}}(t) = 1$. Thus, $\displaystyle \hat{q}_k(\hat{\mu}; 1-\alpha) \leq \max_{1\leq j\leq k} \max_{1\leq i\leq n_j} R_{ji}(\hat{\mu})$, and the length of $C^\modelAvg(x; \alpha)$ is at most $\displaystyle 2\left(\max_{1\leq j\leq k} \max_{1\leq i\leq n_j} R_{ji}(\hat{\mu})\right)$. 
\end{proof}

\section{ADDITIONAL SIMULATIONS}
\label{appendix:sims}

\subsection{Unsupervised Simulations}

\begin{figure}[H]
\centering
\begin{subfigure}{\textwidth}
\centering
\includegraphics[scale=.55]{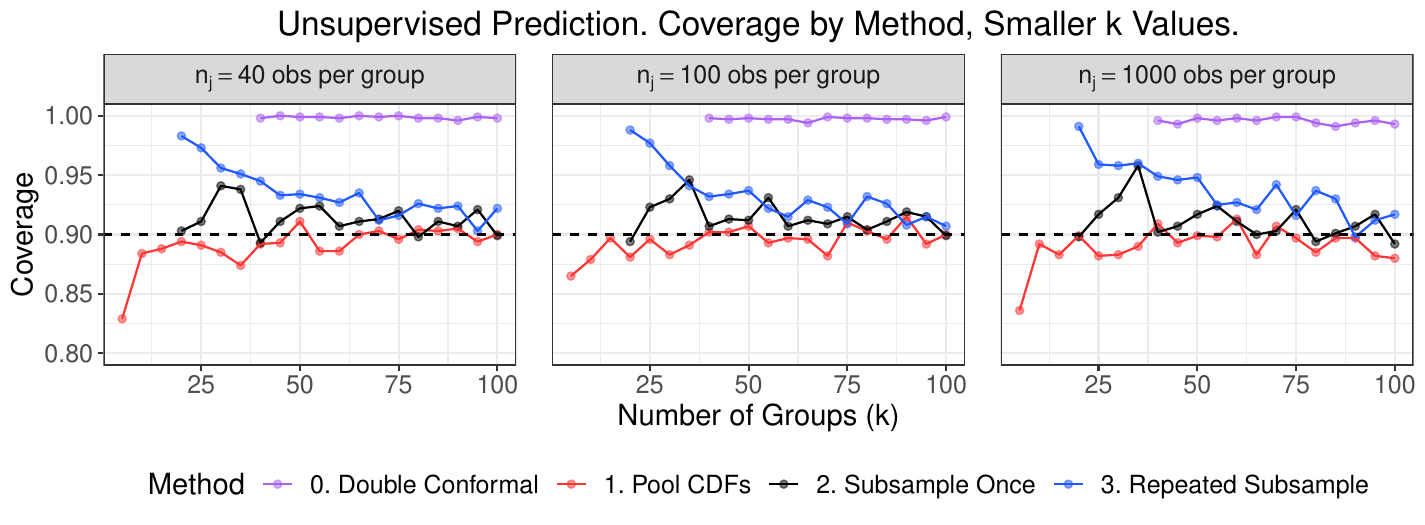}
\subcaption{Smaller numbers of groups ($k$)}
\end{subfigure}
\begin{subfigure}{\textwidth}
\centering
\includegraphics[scale=.55]{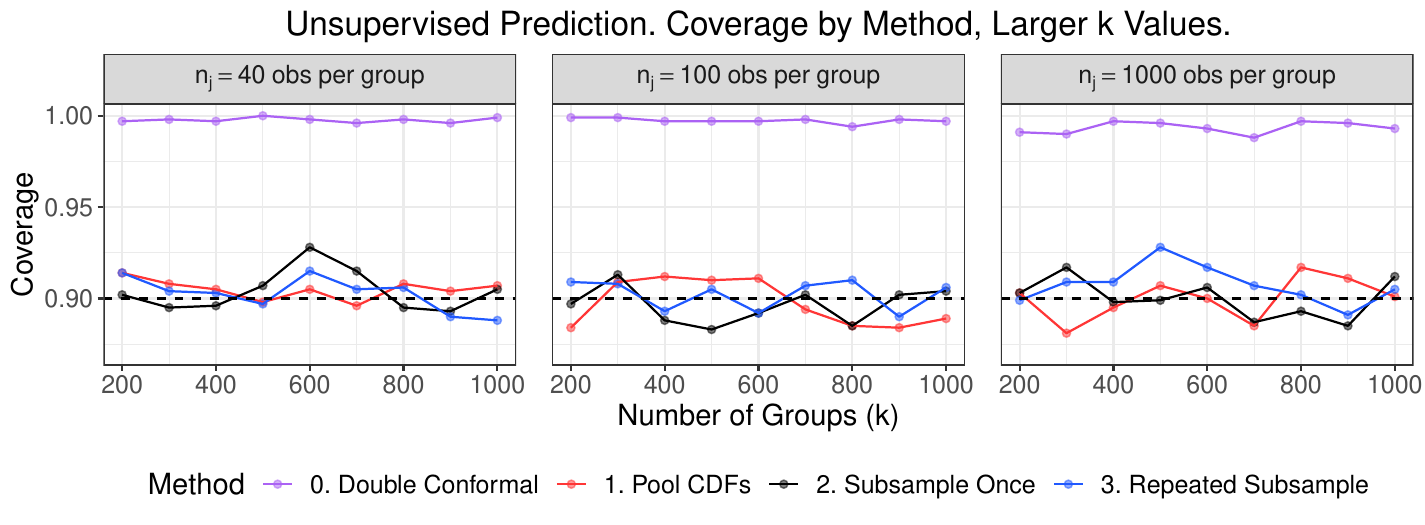}
\subcaption{Larger numbers of groups ($k$)}
\end{subfigure}
\caption{Coverage of unsupervised prediction sets for a new group's observation. Section~\ref{section::simulations_unsup} of the main paper considered $n_j = 100$ observations per group. Setting $n_j = 40$ or $n_j = 1000$ produces similar results.}
\label{fig:coverage_unsup_addl}
\end{figure}

\begin{figure}[H]
\centering
\begin{subfigure}{\textwidth}
\centering
\includegraphics[scale=.55]{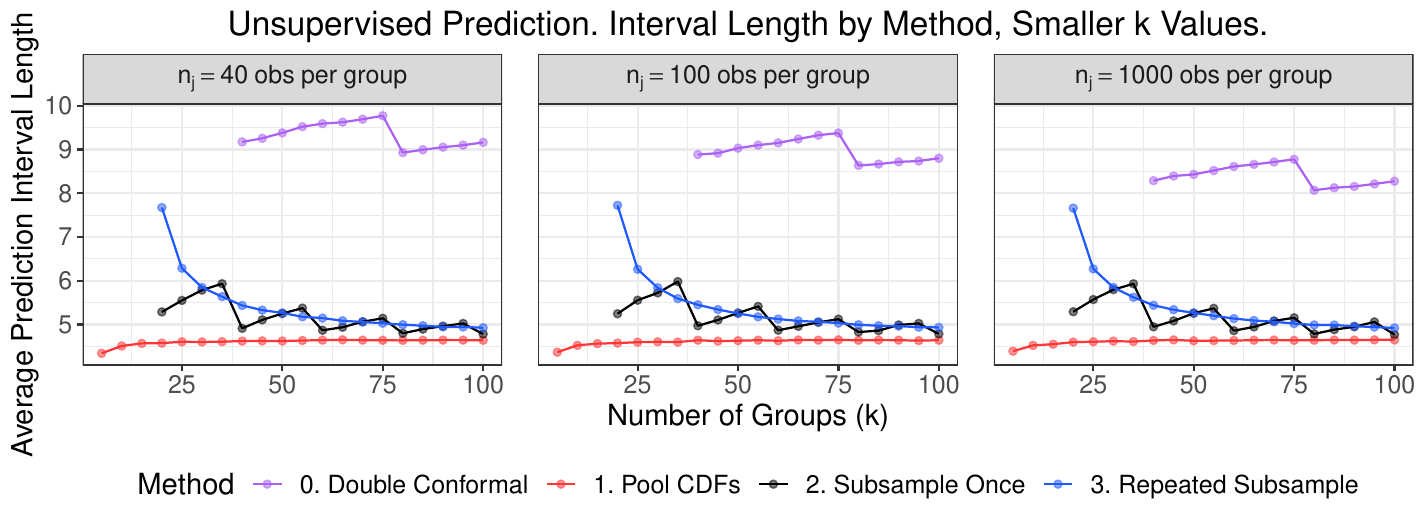}
\subcaption{Smaller numbers of groups ($k$)}
\end{subfigure}
\begin{subfigure}{\textwidth}
\centering
\includegraphics[scale=.55]{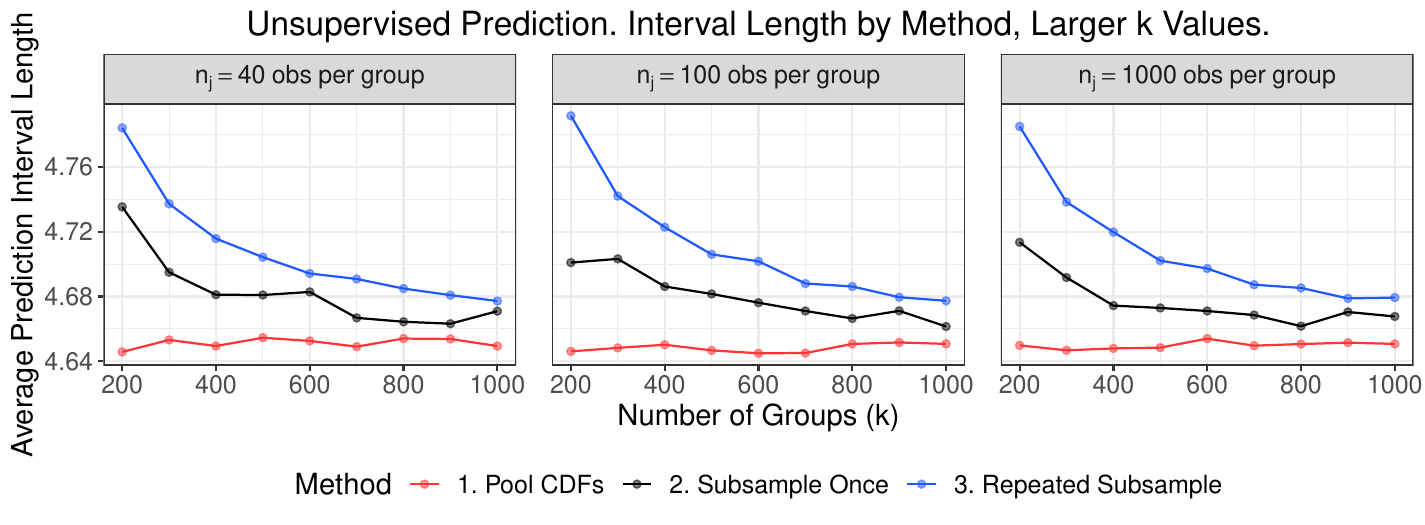}
\subcaption{Larger numbers of groups ($k$)}
\end{subfigure}
\caption{Average unsupervised prediction set length for a new group's observation. Again, $n_j = 40$ and $n_j = 1000$ produce similar results to $n_j = 100$, previously examined in Section~\ref{section::simulations_unsup}.}
\label{fig:size_unsup_addl}
\end{figure}

We consider several additional simulations in the setting of unsupervised prediction on a new distribution. In Section~\ref{section::simulations_unsup}, we considered the following setup: Draw $\theta_1, \ldots, \theta_k \sim N(0,1)$, and simulate $Y_{j1}, \ldots, Y_{jn_j} \sim N(\theta_j, 1)$ for $j = 1,\ldots, k$. We construct prediction intervals for a new $Y \sim N(\theta_{k+1}, 1)$, where $\theta_{k+1}\sim N(0,1)$. Previously, we let $n_j=100$ for all groups. In Figures~\ref{fig:coverage_unsup_addl} and \ref{fig:size_unsup_addl}, we let $n_j = 40$ and $n_j = 1000$ as well. In terms of both coverage and prediction set size, we see very similar results across each value of $n_j$.

\begin{figure}[H]
\centering
\includegraphics[scale=.6]{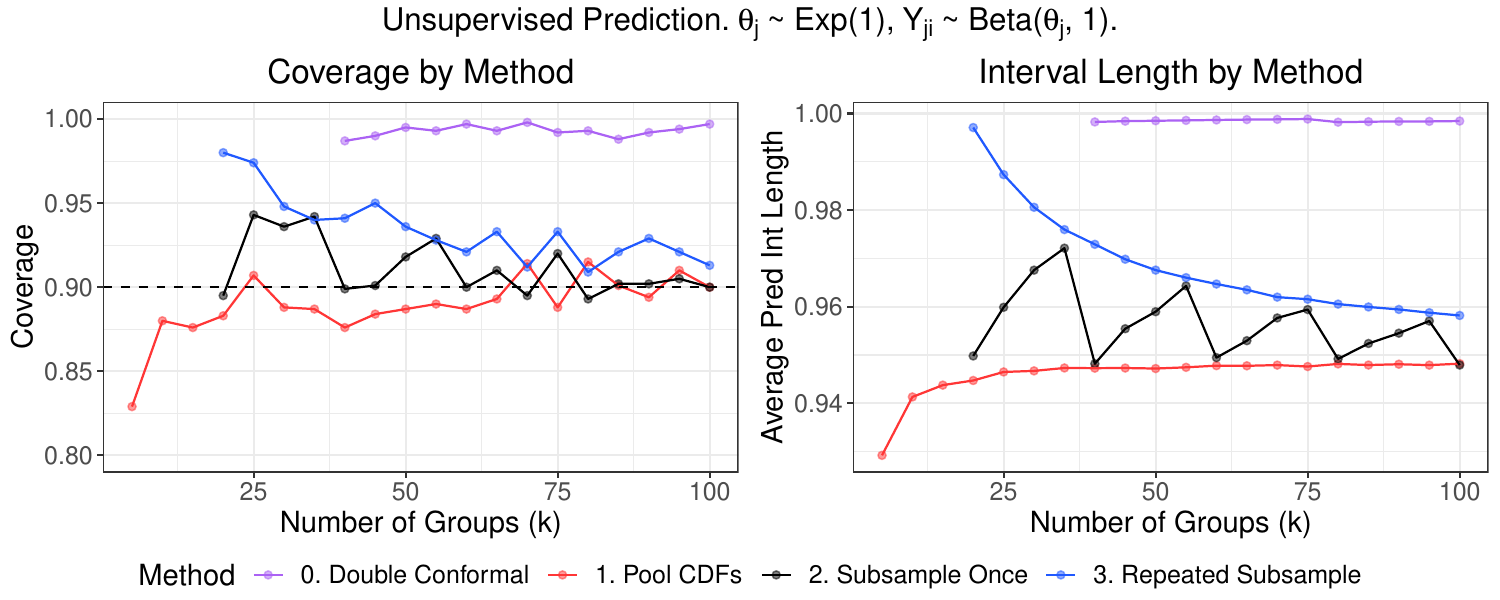}
\caption{Coverage and size of unsupervised prediction sets for a new observation from a new distribution. We simulate $\theta_j \sim \text{Exp}(1)$ and $Y_{ji} \sim \text{Beta}(\theta_j, 1)$, where $j = 1,\ldots, k$ and $i = 1, \ldots, 100$.}
\label{fig:exp_theta_beta_Y}
\end{figure}

\begin{figure}[H]
\centering
\includegraphics[scale=.6]{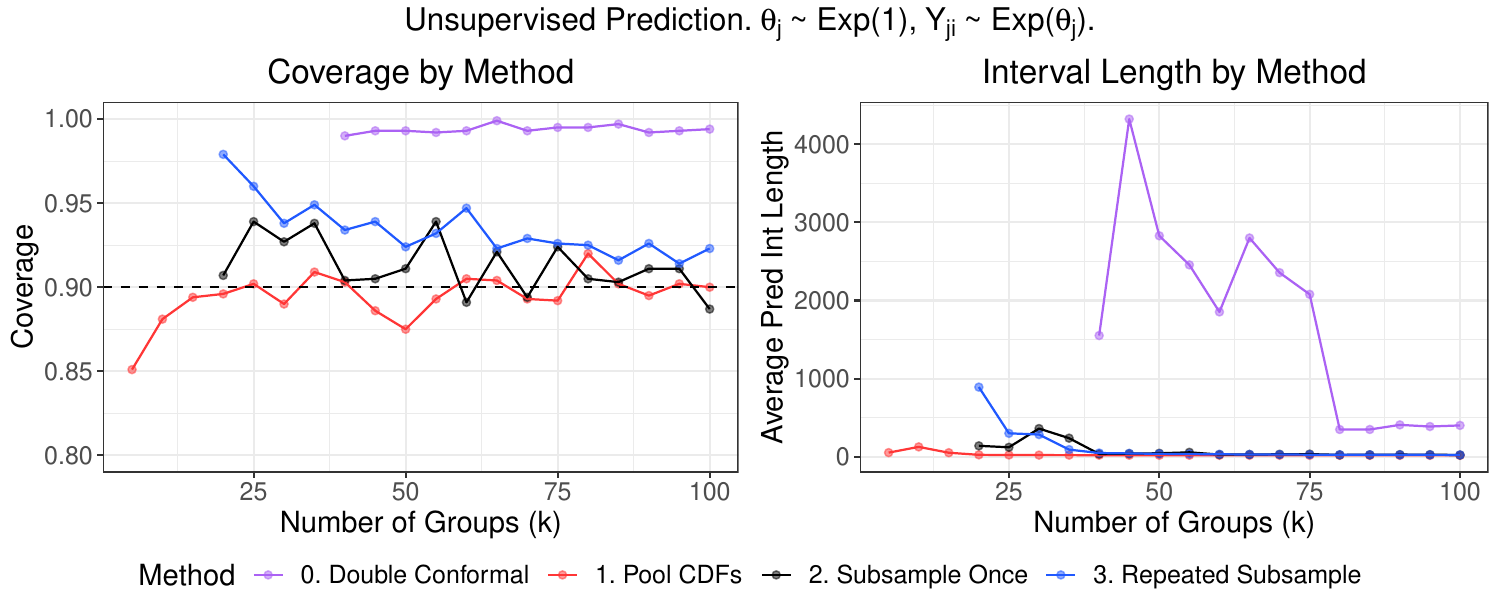}
\caption{Coverage and size of unsupervised prediction sets for a new observation from a new distribution. We simulate $\theta_j \sim \text{Exp}(1)$ and $Y_{ji} \sim \text{Exp}(\theta_j)$, where $j = 1,\ldots, k$ and $i = 1, \ldots, 100$.}
\label{fig:exp_theta_exp_Y}
\end{figure}

We now explore the performance of the unsupervised methods in two settings involving non-normal data. To generate data for Figures~\ref{fig:exp_theta_beta_Y} and \ref{fig:exp_theta_exp_Y}, we draw $\theta_1, \ldots, \theta_k \sim \text{Exp}(1)$. For Figure~\ref{fig:exp_theta_beta_Y}, we simulate $Y_{j1}, \ldots, Y_{jn_j} \sim \text{Beta}(\theta_j, 1)$, $j = 1\ldots, k$. For Figure~\ref{fig:exp_theta_exp_Y}, we simulate $Y_{j1}, \ldots, Y_{jn_j} \sim \text{Exp}(\theta_j)$.  We vary the number of groups ($k$) from 5 to 100 in increments of 5, and we set the number of observations per group ($n_j$) to 100. As in Section~\ref{section::simulations_unsup}, the repeated subsampling sets use $B = 100$ subsamples. We set $\alpha = 0.1$, and we perform 1000 simulations at each $k$. Each simulation generates a data sample, draws a new $Y$ from a new distribution, constructs prediction sets $C(\alpha)$, determines the size of each prediction set, and checks whether $Y\in C(\alpha)$.

Similar to the simulations from Section~\ref{section::simulations_unsup}, double conformal overcovers with coverage of approximately 1. CDF pooling has approximately nominal coverage but often slightly undercovers for small to moderate $k$ (e.g., $k \leq 50$). Single subsampling and repeated subsampling slightly overcover. In line with the coverage, double conformal has the largest prediction intervals, followed by the two subsampling methods. CDF pooling produces the smallest prediction intervals. These simulations align with our recommendations from Section~\ref{section::simulations_unsup}. CDF pooling produces the smallest prediction intervals, with the caveats that it requires continuous $Y$ and often slightly undercovers for small to moderate $k$. If we require methods with theoretical guarantees on the coverage, then the subsampling methods are better choices.

\subsection{Supervised Simulations}

In Section~\ref{section::simulations_sup}, we examined the supervised prediction methods through simulations with the following setup: We draw
\begin{align*}
\theta_1,\ldots,\theta_k &\sim N(\mu,\tau^2) \\
X_{j1},\ldots,X_{jn_j} &\sim N(0,1) \\
\epsilon_{j1},\ldots,\epsilon_{jn_j} &\sim N(0,1).
\end{align*}
We let $Y_{ji} = \theta_j X_{ji} + \epsilon_{ji}$, $j=1,\ldots,k$, $i = 1,\ldots,n_j$. Then we draw a new $X \sim N(0,1)$, $\theta_{k+1}\sim N(\mu, \tau^2)$, and $Y\sim N(\theta_{k+1}X, 1)$. We construct prediction intervals $C(x; \alpha)$ such that $\probnew(Y\in C(X; \alpha)) \geq 1-\alpha$. The simulations in Section~\ref{section::simulations_sup} considered $\mu = 0$, $\tau^2 = 1$, and $n_j = 100$ observations per group. We now consider additional simulations at $(\mu = 0, \tau^2 = 1)$ and $(\mu = 1, \tau^2 = 0.1)$ for $n_j \in \{20, 100, 1000\}$. 
The first pair of $(\mu, \tau^2)$ parameters represents a case where the relationships between $X$ and $Y$ may be quite different across groups. The second pair of $(\mu, \tau^2)$ parameters is a case where the groups have similar trends that relate $X$ and $Y$. 

\begin{figure}[H]
\centering
\begin{subfigure}{\textwidth}
\centering
\includegraphics[scale=.6]{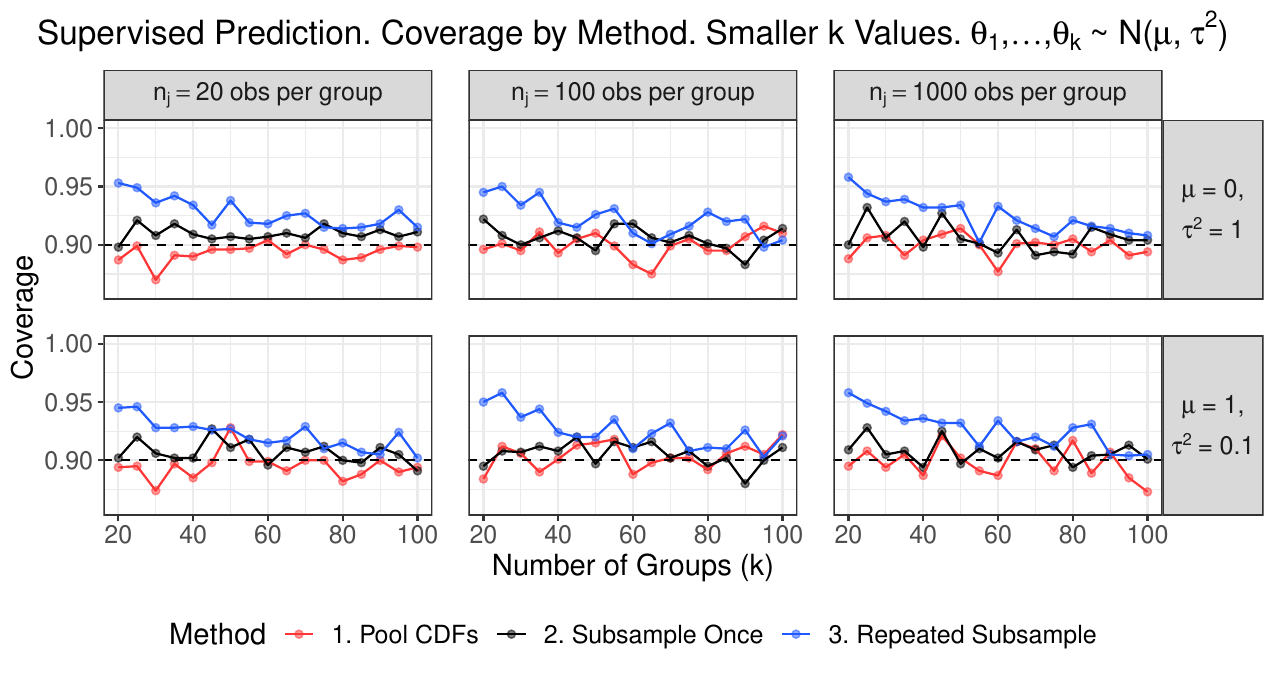}
\subcaption{Smaller numbers of groups ($k$)}
\end{subfigure}
\begin{subfigure}{\textwidth}
\centering
\includegraphics[scale=.6]{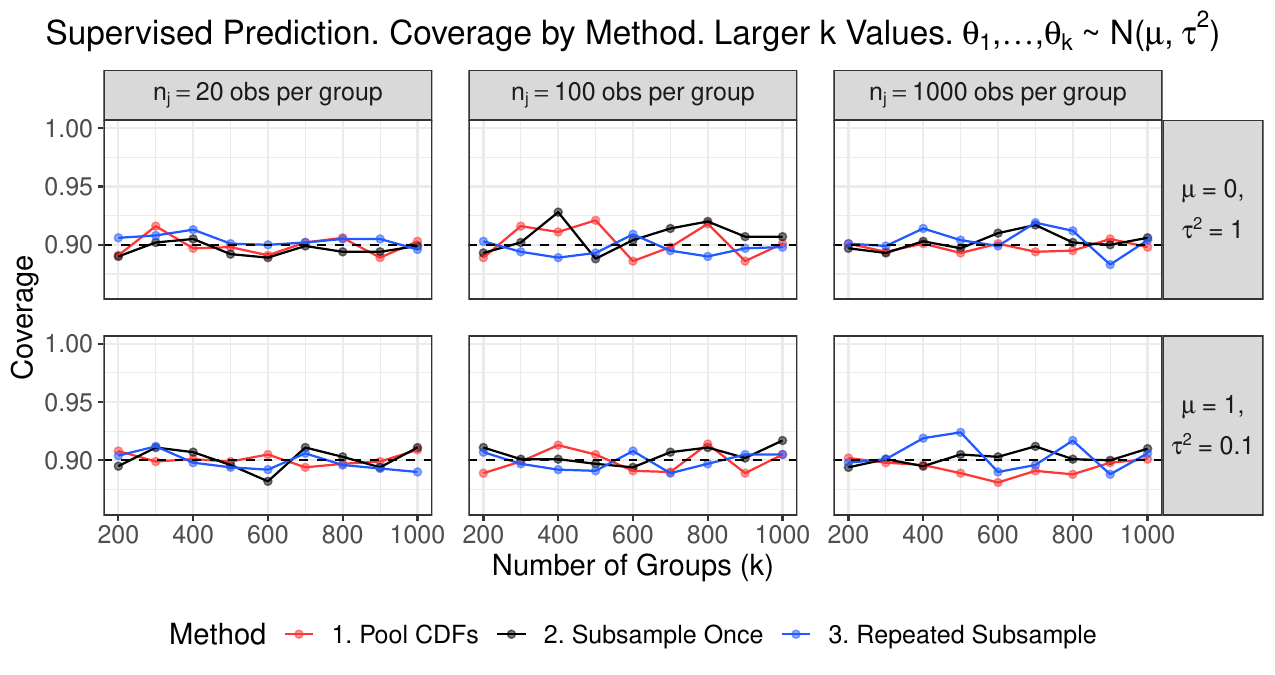}
\subcaption{Larger numbers of groups ($k$)}
\end{subfigure}
\caption{Coverage of supervised prediction sets for a new group's observation. Sample sizes of $n_j \in \{20, 100, 1000\}$ and parameter values of $(\mu = 0, \tau^2 = 1)$ and $(\mu = 1, \tau^2 = 0.1)$ produce similar coverage.}
\label{fig:coverage_sup_addl}
\end{figure}

\begin{figure}[H]
\centering
\begin{subfigure}{\textwidth}
\centering
\includegraphics[scale=.6]{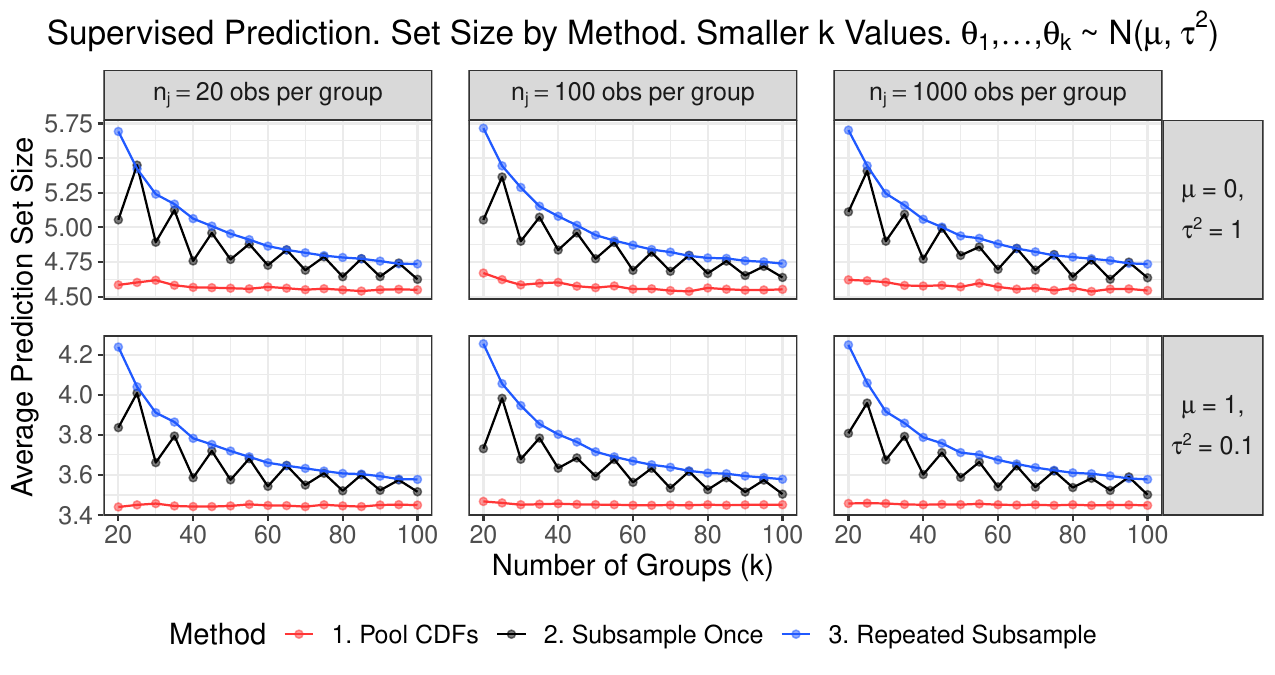}
\subcaption{Smaller numbers of groups ($k$)}
\end{subfigure}
\begin{subfigure}{\textwidth}
\centering
\includegraphics[scale=.6]{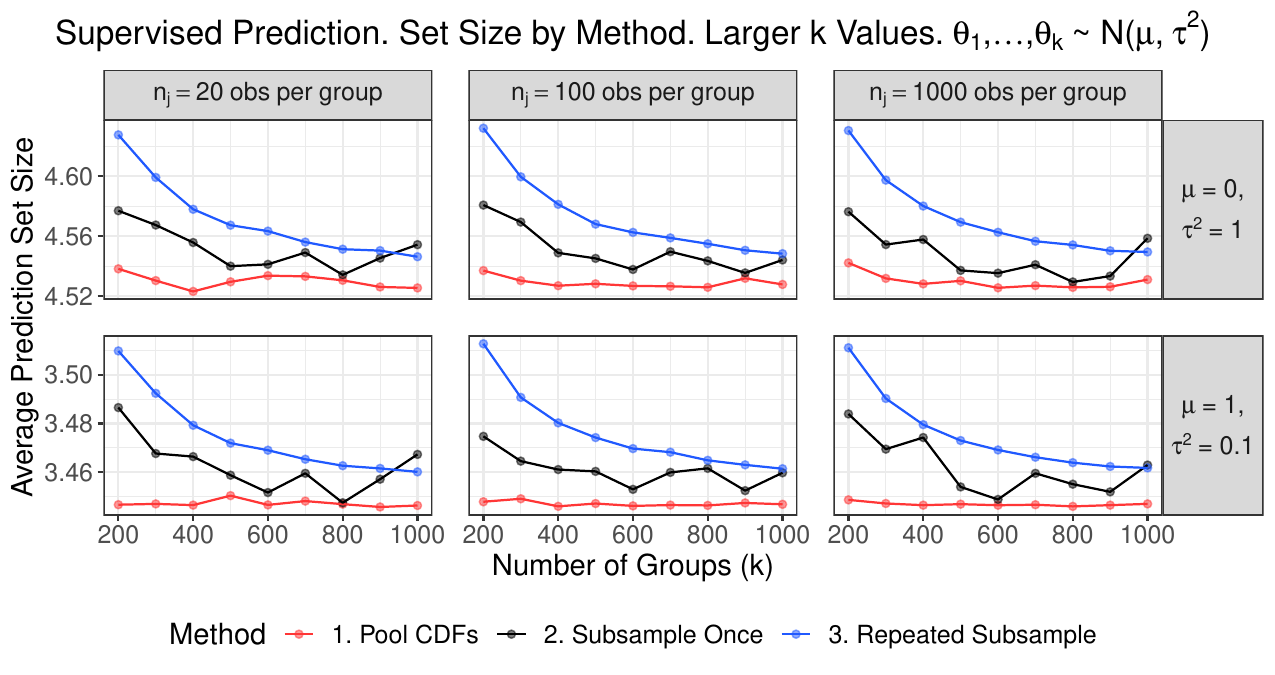}
\subcaption{Larger numbers of groups ($k$)}
\end{subfigure}
\caption{Average supervised prediction set length for a new group's observation. The relationship between the three methods is similar to that of the simulations in Section~\ref{section::simulations_sup}.}
\label{fig:size_sup_addl}
\end{figure}

Figures~\ref{fig:coverage_sup_addl} and \ref{fig:size_sup_addl} show that the coverage and size remain consistent across the choices of $n_j$. The coverage at $(\mu = 0, \tau^2 = 1)$ is similar to the coverage at $(\mu = 1, \tau^2 = 0.1)$. In addition, although the scale of the prediction intervals differs across $(\mu, \tau^2)$ parameters, the relationship between the three methods remains the same. Hence, based on these simulations, we maintain the same recommendations from Section~\ref{section::simulations_sup}.

\end{document}